%% file: main.tex
\documentclass[sigconf]{acmart-primary-2025/acmart}
\AtBeginDocument{%
  }

\copyrightyear{2025}
\acmYear{2025}
\setcopyright{acmlicensed}\acmConference[KDD '25]{Proceedings of the 31st ACM SIGKDD Conference on Knowledge Discovery and Data Mining V.2}{August 3--7, 2025}{Toronto, ON, Canada}
\acmBooktitle{Proceedings of the 31st ACM SIGKDD Conference on Knowledge Discovery and Data Mining V.2 (KDD '25), August 3--7, 2025, Toronto, ON, Canada}
\acmDOI{10.1145/3711896.3737437}
\acmISBN{979-8-4007-1454-2/2025/08}


\usepackage{graphicx}
\usepackage{float} 
\usepackage{enumitem}
\usepackage{xcolor}
\usepackage{hyperref}
\usepackage{listings}
\usepackage{float}
\usepackage{subfigure}
\usepackage{pifont}
\usepackage{xspace}
\usepackage{multirow}

\begin{document}

\title{A Guide to Misinformation Detection Data and Evaluation}



\author{Camille Thibault}
\authornote{Authors contributed equally to this research.}
\affiliation{%
  \institution{Mila, Université de Montréal}
}

\author{Jacob-Junqi Tian}
\authornotemark[1]
\affiliation{%
  \institution{Mila, Vector Institute}
}

\author{Gabrielle Peloquin-Skulski}
\affiliation{%
  \institution{MIT}
  }

\author{Taylor Lynn Curtis}
\affiliation{%
  \institution{Mila}
}

\author{James Zhou}
\affiliation{%
   \institution{UC Berkeley}
}

\author{Florence Laflamme}
 \affiliation{%
   \institution{Mila, Université de Montréal}
 }

\author{Luke Yuxiang Guan}
 \affiliation{%
   \institution{McMaster University}
 }

\author{Reihaneh Rabbany}
 \affiliation{%
   \institution{Mila, McGill University}
 }

\author{Jean-François Godbout}
 \affiliation{%
   \institution{Mila, Université de Montréal}
 }

\author{Kellin Pelrine}
 \affiliation{%
   \institution{Mila, McGill University}
 }
\email{kellin.pelrine@mila.quebec}
\renewcommand{\shortauthors}{Camille Thibault et al.}

\newcommand{\brandname}{CDL-M}
\newcommand\missingmacro{\xspace--\xspace}


\begin{abstract}

\input{000abstract}
\end{abstract}




\makeatletter
\def\@ACM@checkaffil{
    \if@ACM@instpresent\else
    \ClassWarningNoLine{\@classname}{No institution present for an affiliation}%
    \fi
    \if@ACM@citypresent\else
    \ClassWarningNoLine{\@classname}{No city present for an affiliation}%
    \fi
    \if@ACM@countrypresent\else
        \ClassWarningNoLine{\@classname}{No country present for an affiliation}%
    \fi
}
\makeatother

\maketitle

\input{010introduction}
\input{031abbreviated_collection}
\input{040data_quality}
\input{050detection_eval}

\input{060conclusion}

\bibliography{main}
\bibliographystyle{ACM-Reference-Format}

\appendix
\input{070appendix}

\end{document}

%% file: 000abstract.tex
Misinformation is a complex societal issue, and mitigating solutions are difficult to create due to data deficiencies. To address this, we have curated the largest collection of (mis)information datasets in the literature, totaling 75. From these, we evaluated the quality of 36 datasets that consist of statements or claims, as well as the 9 datasets that consist of data in purely paragraph form. We assess these datasets to identify those with solid foundations for empirical work and those with flaws that could result in misleading and non-generalizable results, such as spurious correlations, or examples that are ambiguous or otherwise impossible to assess for veracity. We find the latter issue is particularly severe and affects most datasets in the literature. We further provide state-of-the-art baselines on all these datasets, but show that regardless of label quality, categorical labels may no longer give an accurate evaluation of detection model performance. Finally, we propose and highlight Evaluation Quality Assurance (EQA) as a tool to guide the field toward systemic solutions rather than inadvertently propagating issues in evaluation. Overall, this guide aims to provide a roadmap for higher quality data and better grounded evaluations, ultimately improving research in misinformation detection. All datasets and other artifacts are available at \href{https://misinfo-datasets.complexdatalab.com/}{misinfo-datasets.complexdatalab.com}.

%% file: 010introduction.tex
\section{Introduction}
\label{sec:intro}

Misinformation is a pressing concern for society, already causing significant negative impacts and posing even greater risks with the advent of generative AI \citep{torkington2024ai}. Extensive research has been devoted to this problem, yet it remains unresolved. There has been considerable recent progress in methods, especially leveraging LLMs to detect false information at scale \citep{chen2023combating}. However, to fuel further progress, we also need strong and reliable data.

Multiple studies have identified data availability, and especially data quality, as a barrier for reliable misinformation detection. To begin, obtaining high quality veracity labels is challenging and time-consuming, even for experts \citep{Zubiaga2016}. Shortcuts, though, can cause severe spurious correlations \citep{pelrine2021surprising, wu2022probing}, and even with high quality labels there can be issues with ambiguity of input texts \citep{pelrine2023towards}. While several surveys have mapped the methodological landscape in this domain \citep{shu2017fake, oshikawa2018survey, zhou2020survey, chen2023combating}, the analysis of dataset quality remains either limited in scale \citep{pelrine2021surprising, wu2022probing, pelrine2023towards} or lacking in depth.

To overcome this problem, we present the largest-scale survey in the literature to date, curating 75 datasets with comprehensive descriptive analyses and categorizations. This is nearly three times as many as other dataset-focused surveys like \cite{hamed2023review}, and many times more than general surveys like those of \cite{ali2022fake,shu2017fake,oshikawa2018survey,zhou2020survey}. We provide a summary of each dataset, along with key information like topic, size, modality, languages, geographic region, and time period. 

We further focus on 36 datasets that include claims and 9 datasets that include paragraphs, and evaluate their quality in depth. First, we examine two types of potential spurious correlations that could lead to predictions based on invalid, non-generalizable signals. In particular, we start by looking at keyword based correlations before following up with temporal correlations. Both of these can serve as proxies for topics, events, and other correlated information. We then assess whether the examples in these datasets are actually feasible to assess for veracity at all. We find that most datasets contain substantial ambiguities and other issues, such that over half of the claims data may be infeasible for methods without evidence retrieval, with a large portion remaining infeasible even with retrieval. These quality assessments offer both immediately practical insights for selecting the most reliable datasets, and significant implications for directions in the field of misinformation research overall.

Our work also addresses challenges of model evaluation after researchers complete the dataset selection phase. We first present a unified formatting and labeling schema for all 36 claims datasets and 9 paragraph datasets. Next, we establish state-of-the-art baselines using GPT-4 with and without web search to collect evidence for veracity. Following this analysis, we find that standard evaluation metrics like accuracy and F1, when computed simply in relation to ground truth labels, are no longer sufficient to evaluate leading generative methods for misinformation detection and could lead to invalid conclusions. To address this challenge, we present and test an alternative evaluation approach, emphasizing the need for future research in this domain.

\newcommand{\cdlmd}{CDL-MD\xspace} 
\newcommand{\cdlme}{CDL-DQA\xspace} 
\newcommand{\cdlma}{CDL-EQA\xspace} 

Finally, our findings come together in our recommendation for a new practice of \textbf{Evaluation Quality Assurance (EQA)}, which offers a tool to improve data selection and evaluation methodologies. In summary, we present a guide to misinformation datasets, including:
\begin{itemize}[left=10pt,topsep=2pt,noitemsep]
    \item The largest scale collection of misinformation datasets, \textbf{\cdlmd}, with a unified labeling schema, made easily accessible through a HuggingFace Repository; along with baseline performances on these datasets.
    \item An essential toolkit, called \textbf{\cdlme}, for quality evaluation of misinformation datasets to analyze spurious correlations as well as the feasibility of the datapoints. Applying this toolkit to \cdlmd reveals numerous quality issues (summarized in Table~\ref{tab:summary_quality}), such as how the majority of claims datasets contain numerous examples whose veracity may be impossible to evaluate.
    \item Recommended practices of \textbf{EQA}, for research proposing misinformation detection methods. Informed by our findings of severe quality issues in both data and evaluation procedures, such as how simple metrics like accuracy and F1 may be obsolete in this domain, we propose that assurance of evaluation quality is an indispensable component for future research here.
\end{itemize}

We provide links to our unified dataset collection \cdlmd on HuggingFace, our tools \cdlme and other code, and other outputs through our website \href{https://misinfo-datasets.complexdatalab.com/}{misinfo-datasets.complexdatalab.com}.

\input{tables/summary_final_evaluation}

\begin{figure*}[ht]
    \centering
    \begin{minipage}{0.3\textwidth}
        \centering
        \includegraphics[trim={0cm 1.0cm 1.4cm 0},clip,width=\textwidth]{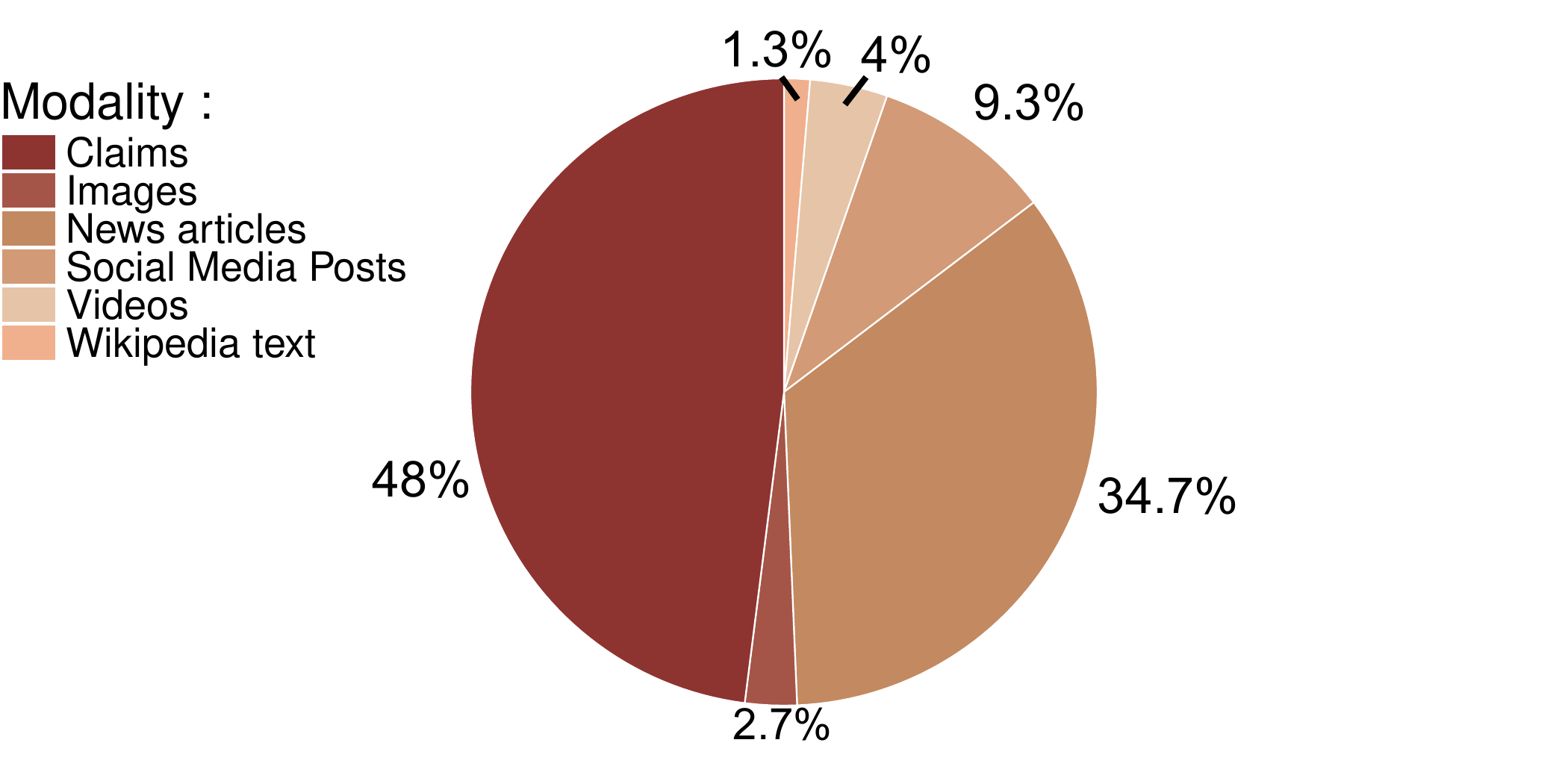}
        \label{fig:modality_graph}
    \end{minipage}
    \hspace{2mm} 
    \begin{minipage}{0.33\textwidth}
        \centering
        \includegraphics[trim={.4cm .9cm 2cm 0},clip,width=\textwidth]{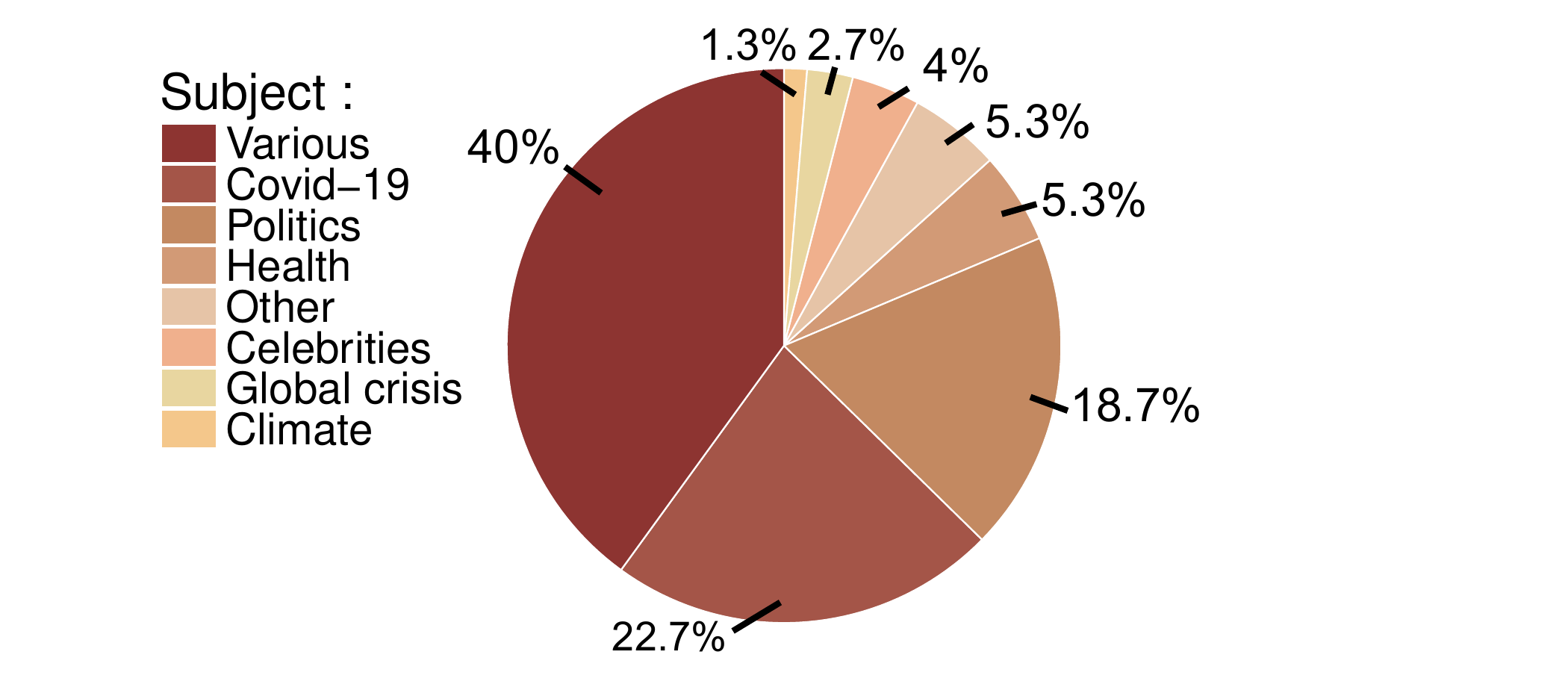}
        \label{fig:subject_graph}
    \end{minipage}%
    \hspace{2mm} 
    \begin{minipage}{0.33\textwidth}
        \centering
        \includegraphics[trim={.4cm 1.0cm 2.5cm 0},clip,width=\textwidth]{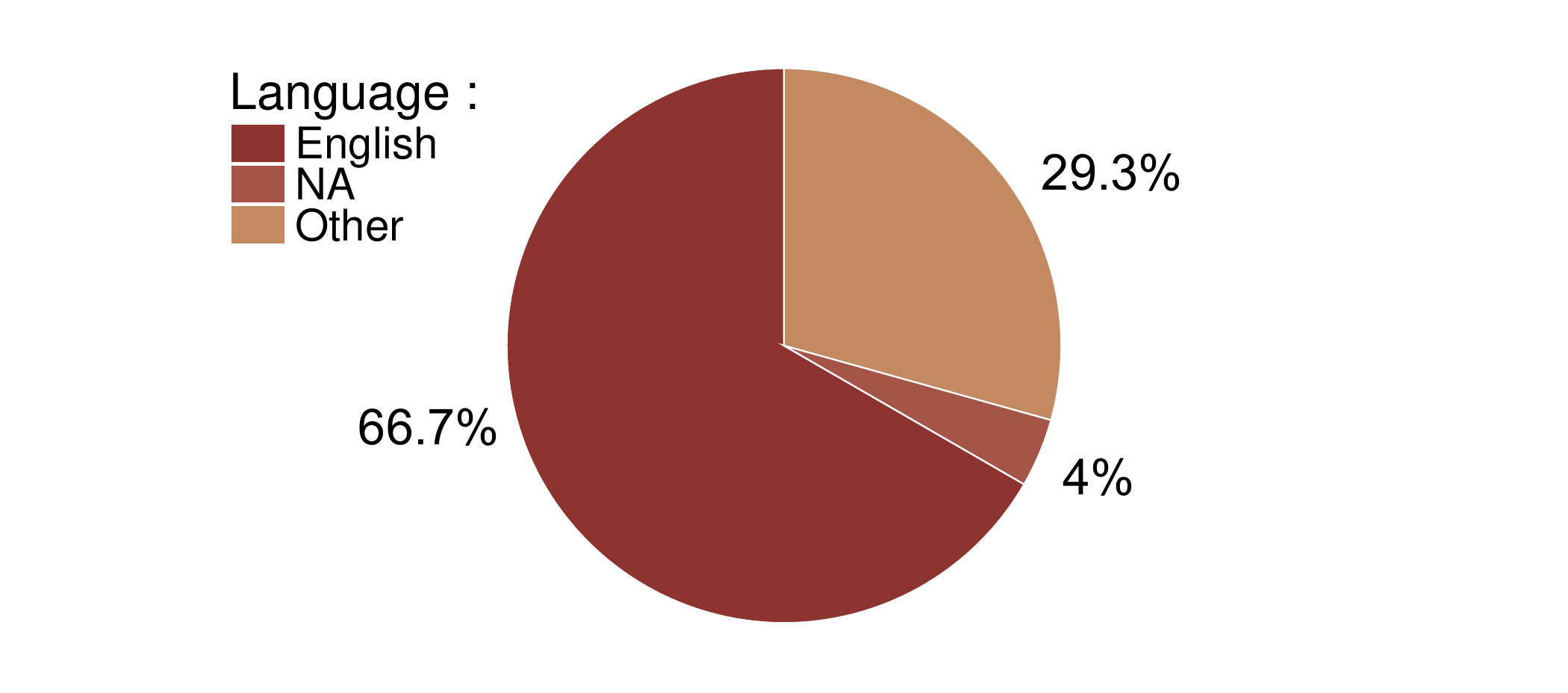}
        \label{fig:language_graph}
    \end{minipage}
    \caption{An overview of the modalities, subject counts, and languages of datasets in \cdlmd. For more details on these and other attributes like size, geographic region, and date of the data, please refer to the comprehensive table in Appendix \ref{app:datacollection}.}
    \label{fig:aligned_graphs}
\end{figure*}

%% file: tables/summary_final_evaluation.tex
\begin{table}[h]
\caption{Dataset quality assessments. A \textcolor[rgb]{0,0.6,0}{\checkmark} denotes a dataset that passes the evaluation criterion. The horizontal line divides claim and paragraph datasets.\vspace{1mm}\\\textcolor{red}{\ding{55}} in Keyword or Temporal indicates supervised methods may learn significant spurious correlations of the respective type. \\ \textcolor{red}{\ding{55}} in Feasibility indicates over a quarter of the data may be impossible to assess for veracity at all.}
\label{tab:summary_quality}
\centering
\fontsize{9}{11}\selectfont
\resizebox{\linewidth}{!}{
\begin{tabular}[t]{lccc}
\toprule
Dataset & Keywords & Temporal & Feasibility \\
\midrule
AntiVax & - & \textcolor[rgb]{0,0.6,0}{\checkmark} & - \\
Check-COVID & \textcolor[rgb]{0,0.6,0}{\checkmark} & - &  \textcolor{red}{\ding{55}}\\
ClaimsKG & \textcolor[rgb]{0,0.6,0}{\checkmark} & \textcolor{red}{\ding{55}} &  \textcolor{red}{\ding{55}}\\ 
Climate-Fever & \textcolor[rgb]{0,0.6,0}{\checkmark} & - & \textcolor[rgb]{0,0.6,0}{\checkmark}\\
CMU-MisCOV19 & - & \textcolor[rgb]{0,0.6,0}{\checkmark} & - \\
CoAID & \textcolor{red}{\ding{55}} & \textcolor[rgb]{0,0.6,0}{\checkmark} &  \textcolor{red}{\ding{55}}\\
COVID-19-Rumor & \textcolor[rgb]{0,0.6,0}{\checkmark} & - &  \textcolor{red}{\ding{55}}\\
Covid-19-disinformation & - & \textcolor[rgb]{0,0.6,0}{\checkmark} & - \\
COVID-Fact & \textcolor[rgb]{0,0.6,0}{\checkmark} & - &  \textcolor{red}{\ding{55}}\\
DeFaktS & \textcolor[rgb]{0,0.6,0}{\checkmark} & \textcolor[rgb]{0,0.6,0}{\checkmark} &  \textcolor{red}{\ding{55}} \\
ESOC Covid-19 & - & - &  \textcolor{red}{\ding{55}} \\
FakeCovid & \textcolor[rgb]{0,0.6,0}{\checkmark} & \textcolor[rgb]{0,0.6,0}{\checkmark} &  \textcolor{red}{\ding{55}}\\
FaVIQ & \textcolor[rgb]{0,0.6,0}{\checkmark} & - & \textcolor[rgb]{0,0.6,0}{\checkmark}\\
FEVER & \textcolor[rgb]{0,0.6,0}{\checkmark} & - & \textcolor[rgb]{0,0.6,0}{\checkmark}\\
FEVEROUS & \textcolor[rgb]{0,0.6,0}{\checkmark} & - & \textcolor[rgb]{0,0.6,0}{\checkmark}\\
FibVID & \textcolor[rgb]{0,0.6,0}{\checkmark} & \textcolor{red}{\ding{55}} &  \textcolor{red}{\ding{55}}\\
HoVer & \textcolor[rgb]{0,0.6,0}{\checkmark} & - & \textcolor[rgb]{0,0.6,0}{\checkmark}\\
IFND & \textcolor{red}{\ding{55}} & - &  \textcolor{red}{\ding{55}} \\
LIAR & \textcolor[rgb]{0,0.6,0}{\checkmark} & - &  \textcolor{red}{\ding{55}}\\
LIAR-New & \textcolor[rgb]{0,0.6,0}{\checkmark} & \textcolor[rgb]{0,0.6,0}{\checkmark} &  \textcolor{red}{\ding{55}}\\
MediaEval & - & \textcolor{red}{\ding{55}} & - \\
MM-COVID & \textcolor{red}{\ding{55}} & - &  \textcolor{red}{\ding{55}}\\
MultiClaim & - & - &  \textcolor{red}{\ding{55}} \\
NLP4IF-2021 & \textcolor[rgb]{0,0.6,0}{\checkmark} & - &  \textcolor{red}{\ding{55}} \\
PHEME & - & - &  \textcolor{red}{\ding{55}} \\
PubHealthTab & \textcolor[rgb]{0,0.6,0}{\checkmark} & - &  \textcolor{red}{\ding{55}} \\
Rumors & \textcolor[rgb]{0,0.6,0}{\checkmark} & \textcolor{red}{\ding{55}} &  \textcolor{red}{\ding{55}}\\
Snopes Fact-news & \textcolor[rgb]{0,0.6,0}{\checkmark} & - &  \textcolor{red}{\ding{55}}\\
TruthSeeker2023 & \textcolor{red}{\ding{55}} & - &  \textcolor{red}{\ding{55}}\\
Twitter15 & \textcolor{red}{\ding{55}} & \textcolor{red}{\ding{55}} &  \textcolor{red}{\ding{55}}\\
Twitter16 & \textcolor{red}{\ding{55}} & \textcolor{red}{\ding{55}} &  \textcolor{red}{\ding{55}} \\
Verite & \textcolor[rgb]{0,0.6,0}{\checkmark} & - &  \textcolor{red}{\ding{55}}\\
WICO & - & \textcolor[rgb]{0,0.6,0}{\checkmark} & - \\
X-Fact & \textcolor[rgb]{0,0.6,0}{\checkmark} & \textcolor{red}{\ding{55}} &  \textcolor{red}{\ding{55}}\\
\midrule
BanFakeNews & \textcolor{red}{\ding{55}} & \textcolor{red}{\ding{55}} & \textcolor{red}{\ding{55}}\\
BenjaminPoliticalNews & \textcolor{red}{\ding{55}} & - & \textcolor{red}{\ding{55}} \\
Celebrity & \textcolor{red}{\ding{55}} & - & \textcolor[rgb]{0,0.6,0}{\checkmark}\\
CT-FAN & \textcolor{red}{\ding{55}} & - & \textcolor[rgb]{0,0.6,0}{\checkmark}\\
FA-KES & \textcolor[rgb]{0,0.6,0}{\checkmark} & \textcolor[rgb]{0,0.6,0}{\checkmark} & \textcolor[rgb]{0,0.6,0}{\checkmark}\\
FakeNewsAMT & \textcolor[rgb]{0,0.6,0}{\checkmark} & - & \textcolor[rgb]{0,0.6,0}{\checkmark}\\
FakeNewsCorpus & - & - & \textcolor{red}{\ding{55}} \\
ISOT Fake News &  \textcolor{red}{\ding{55}} & \textcolor{red}{\ding{55}} & \textcolor[rgb]{0,0.6,0}{\checkmark}\\
TI-CNN &  \textcolor{red}{\ding{55}} & \textcolor{red}{\ding{55}} & \textcolor[rgb]{0,0.6,0}{\checkmark} \\
\bottomrule

\end{tabular}}
\end{table}

%% file: 031abbreviated_collection.tex
\section{\cdlmd}

We curate a combined collection of 75 datasets, named \textbf{\cdlmd} (Complex Data Lab Misinfo Datasets). The collection includes 36 specifically focused on claims and statements (defined here as short texts ranging from one to two sentences), and 9 specifically focused on paragraphs (longer than two sentences). The full collection contains a total of 120,901,495 observations, while the subset that we further analyze includes 1,741,146 observations. These data encompass a wide range of topics, modalities, languages, and other characteristics (partially summarized in Figure~\ref{fig:aligned_graphs}, detailed in Appendix~\ref{sec:Survey}).

Harmonizing labels across different datasets is a crucial step to ensure comparability of results and robustness of analyses. Since each study employs its own criteria for classifying veracity (see Appendix \ref{app:datadetails}), in our collection, we use the original labels from the studies to create a more consistent categorical variable across datasets. Specifically, we classify content as true, false, mixed or unknown. Information that is mostly true is classified as \textit{true}, while content that is mostly false is classified as \textit{false}. Finally, ambiguous claims, such as those partially true or false, were classified as \textit{mixed}, while claims that were unproven, unrelated or contained no information about their veracity were coded as \textit{unknown}. 

An extensive treatment of the construction process of this collection, including how we identified the datasets, additional descriptive information for each one, expanded discussion of the labeling approach, and other discussion, is included in Appendix~\ref{sec:Survey}. 

%% file: 040data_quality.tex
\section{Data Quality}
\label{sec:data_quality}

Shortcut learning is a serious barrier to predictive systems working in the real world \citep{geirhos2020shortcut}. In this section, we assess datasets' potential for teaching algorithms spurious keyword and temporal correlations. These are key considerations when training and evaluating supervised methods. We then assess whether the inputs are making sufficiently complete and unambiguous claims for it to be feasible to predict their veracity at all, which is critical to the evaluation of all types of methods. We provide the automated tools that enable this analysis through our website as the Complex Data Lab Dataset Quality Assessment (\textbf{\cdlme}).

\subsection{Spurious Keyword Correlations}

\begin{figure}[h]
    \centering
    \includegraphics[width=\linewidth]{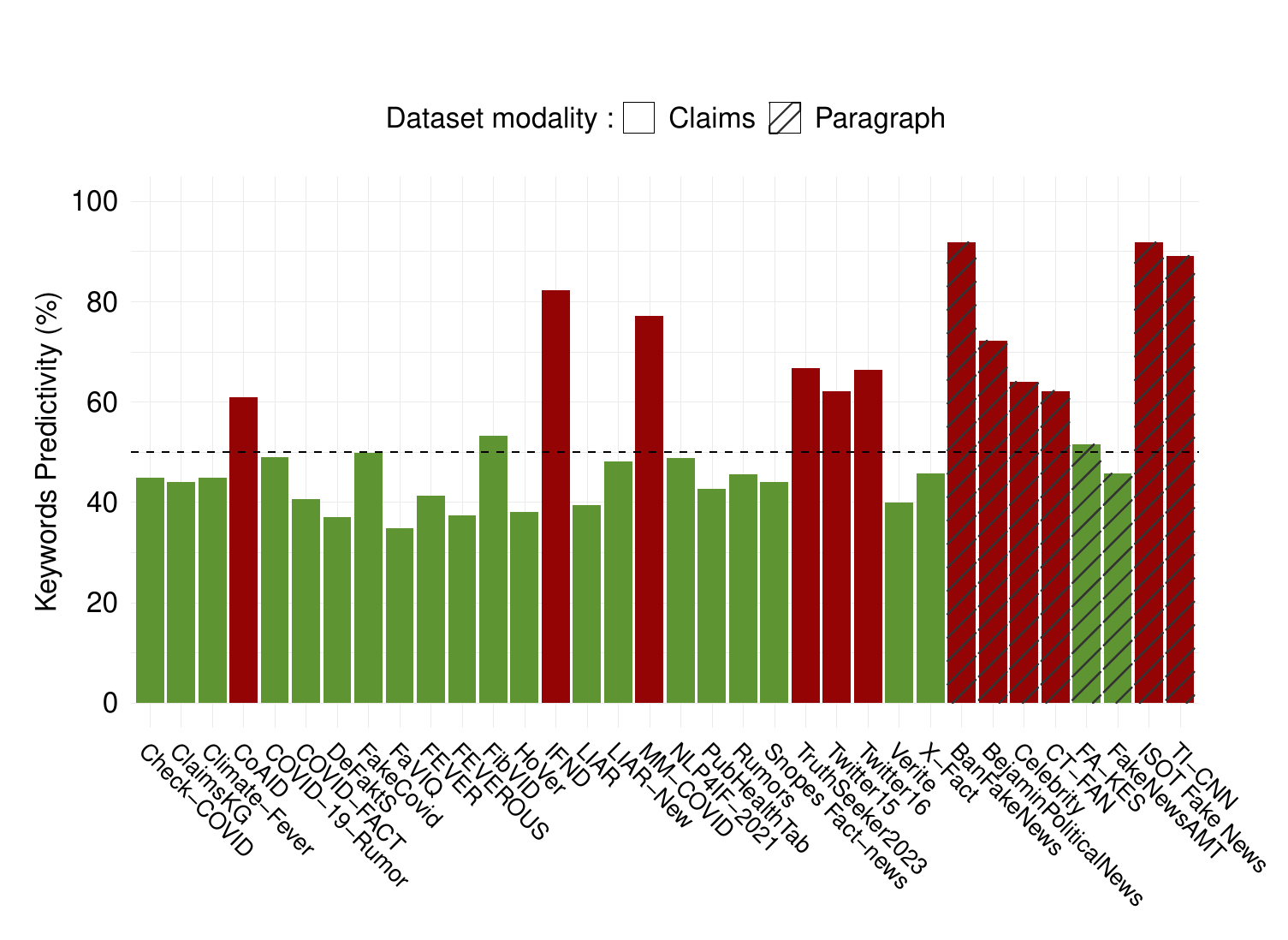} 
    \vspace{-3mm}
    \caption{Keywords correlations evaluation. A high predictivity score that far exceeds the 50\% baseline, indicated by the dashed line, means that the keywords provide an unrealistically strong prediction. Green bars indicate datasets that pass this check, while red bars represent ones with spurious correlations. All numbers are \% macro F1.}
    \label{fig:keywords_graph}
\end{figure}

We first evaluate whether there are certain keywords that overpredict veracity in the datasets. We adapt the approach that \cite{pelrine2021surprising} used to check for spurious temporal correlations. Specifically, we trained a random forest classifier with the 40 most frequent words in each dataset, after removing stop-words and excluding datasets containing only tweet IDs (since in that case the claims are inaccessible), as well as those with single veracity labels that include only false statements. This analysis was conducted in two stages: first by incorporating only the labels True and False, and second, by also including the label Mixed. Utilizing scikit-learn, we set a maximum tree depth of 20 and retained the other default settings. We then compare with a baseline of predicting randomly according to the class label distribution and no other information, providing a reference point for assessing the predictive power of the keywords. Thus, keywords in a dataset are significantly predictive if their performance significantly exceeds the baseline. The macro F1 scores for the true and false labels are presented in Figure \ref{fig:keywords_graph} and Table~\ref{tab:keywords_correlations} provides the results that also incorporate mixed labels and baselines.

We particularly flag six claims datasets for spurious correlations between certain words and labels, as well as six datasets containing paragraphs. Respectively, these are: CoAID, IFND, MM-COVID, TruthSeeker2023, Twitter15 and Twitter16; and BanFakeNews, BenjaminPoliticalNews, Celebrity, CT-FAN, ISOT Fake News and TI-CNN. For example, consider Truthseeker2023. Nearly all tweets here mentioning politicians are labeled as ``false'', with only those containing ``Trump'' showing more variation (see also Appendix~\ref{app:keywords}). Obviously, in the real world, tweets mentioning politicians are not exclusively false. Thus, models trained on data like Truthseeker2023 risk generalizing inaccurate results, and doing so on topics extremely sensitive to bias like discussion of politicians. Moreover, these findings extend beyond political names. Terms like ``michigan'', ``vaccines'', ``immunity'', ``pfizer'', and so on are consistently labeled as false, while words like ``marijuana'', ``wealth'', ``terrorism'', ``radical'' and others are always associated with the true label (see also Figure ~\ref{fig:truthseeker2023} in the Appendix). Similar patterns are also observed in the other datasets with highest keyword predictivity. Therefore, we urge caution about training and testing models on these datasets, especially text-focused models.

\subsection{Spurious Temporal Correlations}

\citet{pelrine2021surprising} highlighted how collecting data of different classes at different times can make temporal information unrealistically predictive. For example, discussion of particular news events can become excessively correlated with veracity labels, leading to artificially inflated performance for classifiers that rely on these events, that will not generalize to the real world where veracity cannot be determined by event or topic alone.

Like in the preceding section, we assess this limitation by training a random forest classifier. As feature, we use either the first three digits of the tweet ID (which contain time information) as in \cite{pelrine2021surprising} for Tweet datasets, or the date itself for datasets which include it. For the latter, we encode it as the integer number of days since the first date in the dataset. We exclude from this analysis datasets without either form of temporal information.

\begin{figure}[h]
    \centering
    \includegraphics[width=\linewidth]{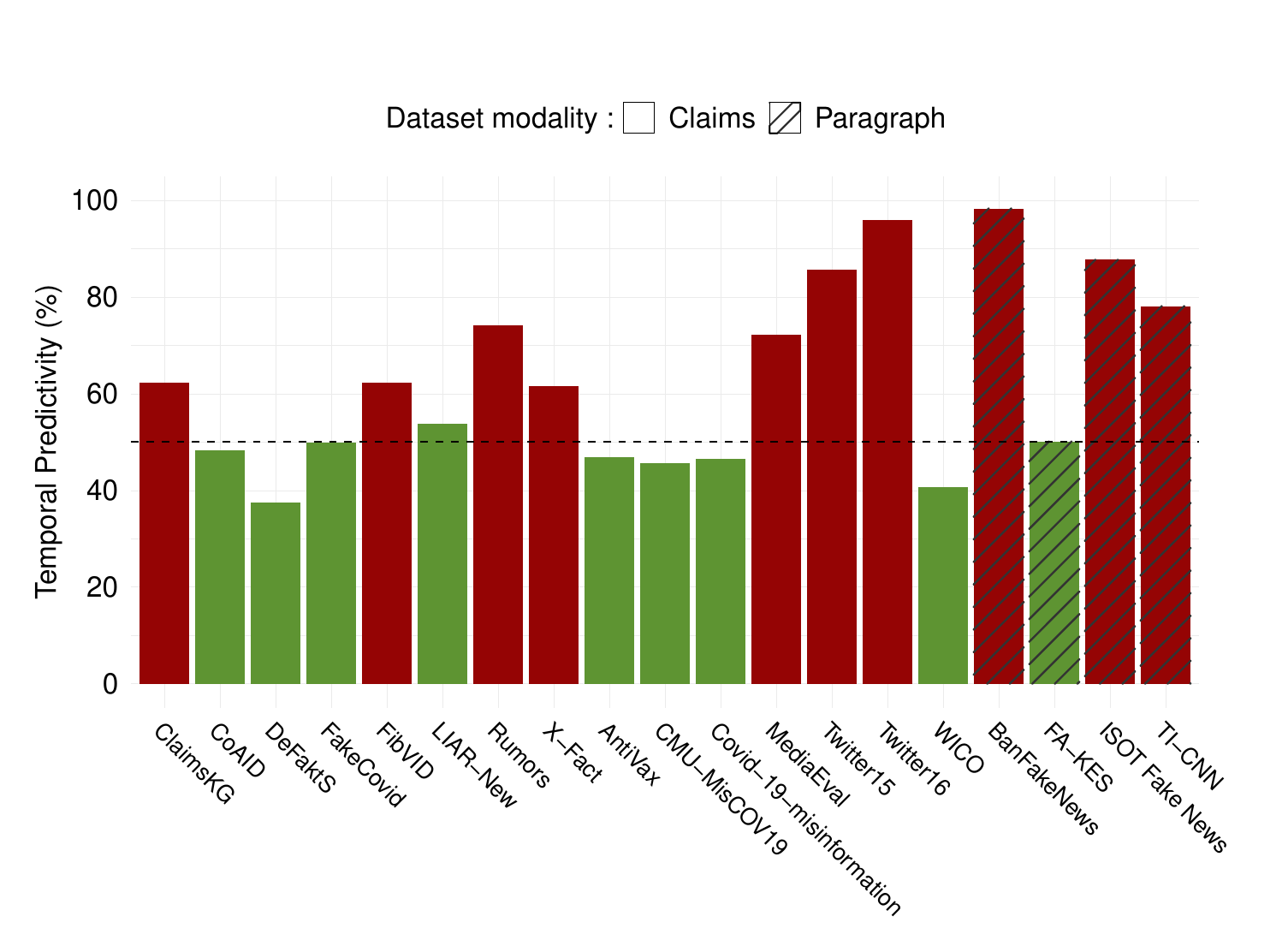} 
    
    \vspace{-2mm}
    \caption{Temporal correlation evaluation. A high score that far exceeds the 50\% baseline means time---and information correlated with it---is unrealistically predictive. Green bars indicate datasets that pass this check, while red bars represent ones with spurious correlations. All numbers are \% macro F1.}
    \label{fig:temporal_graph}
\end{figure}

Results are shown in Figure~\ref{fig:temporal_graph} and Table~\ref{tab:temporal}. We first note that our findings on Twitter15 and Twitter16 are similar to \cite{pelrine2021surprising}, confirming these datasets have extreme issues with spurious correlations in temporal information. They should not be used without carefully and explicitly addressing this limitation. Similarly, the paragraph datasets BanFakeNews, IsotFakeNews, and TI-CNN show high levels of spurious temporal leakage. While not quite as severe, we also see that MediaEval and Rumors also suffer from some significant spurious temporal correlations, and caution is advised. The rest of the datasets have a substantially better temporal balance, with the temporal feature offering little better than random performance. However, we note that only a small fraction of the total datasets include dates, and recommend that future datasets add this important information.

\subsection{Feasibility}

If a claim is too ambiguous, it may be impossible and meaningless---or even misleading---to assess its veracity, irrespective of the power of one's assessment system. For example, it is impossible to evaluate the veracity of the claim ``The senator said the earth is flat'' without knowing which senator is referred to. If a dataset contains too many such examples, it will be problematic to train and evaluate algorithms on it. \cite{pelrine2023towards} performed analysis on a limited, manual scale and found problems of this type in the LIAR dataset. To extend this analysis to the field at large, 8 human expert annotators each labeled examples from 29 datasets. We also developed an AI annotator that gives a comparable assessment, providing both additional evidence and scalability. 

Our annotators categorized the claim text according to the following schema:
\begin{itemize}[left=10pt,topsep=2pt,noitemsep]
    \item \textbf{Feasible}: The statement provides enough context for an AI to determine its truthfulness with certainty.
    \item \textbf{Feasible with web search}: Some key information is missing, preventing an AI from determining the claim’s truthfulness without retrieving additional data online.
    \item \textbf{Not feasible}: The statement is too vague or incomplete for a web search to provide sufficient evidence for verification.
\end{itemize}

This was done for 1230 human-annotated claims dataset examples (uniformly distributed for approximately 42 per dataset) with most annotated by at least two annotators, and 8700 AI-annotated claims examples plus 2700 paragraph examples (300 per dataset). Full human annotator instructions and sampling procedures, LLM prompting procedures, and other supplementary information are available in Appendix~\ref{app:feasibility}.

\begin{figure}[h!]
    \centering
    \includegraphics[width=1\linewidth]{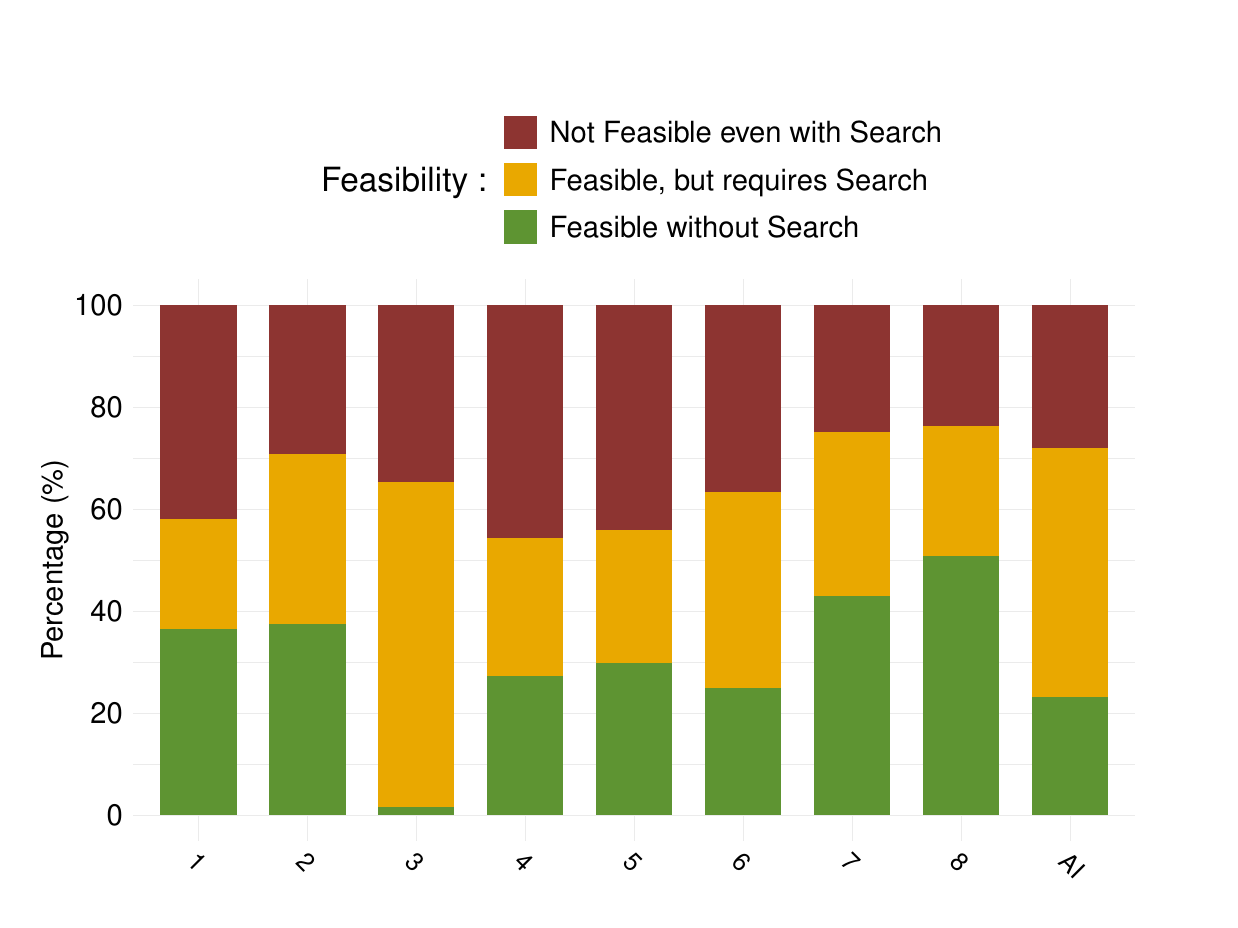}
    \caption{Feasibility assessed by 8 human expert and 1 AI annotators, averaged over all datasets. Without an evidence retrieval system (e.g., web search), most data is not feasible to assess for veracity.}
    \label{fig:feasibility_agg}
\vspace{-3mm}
\end{figure}

We first aggregate over datasets, presenting the individual results of our 9 human and AI annotators in Figure~\ref{fig:feasibility_agg}. Our results show universal agreement that \textbf{without an evidence retrieval (search) system, at least half of claims in these 29 datasets cannot be validly assessed for veracity}. This suggests that such methods are often being evaluated on impossible tasks, and there is a severe risk that evaluation comparing such methods is determining not the best generalizable predictor but the best shortcut learner. We recommend that any proposed methods in this category conduct stringent Evaluation Quality Assurance (see Section~\ref{sec:eqa}), and suggest that applications requiring even slightly out of distribution predictions should ground them in evidence.

\begin{figure*}[htbp]
        \centering
        \includegraphics[width=\textwidth]{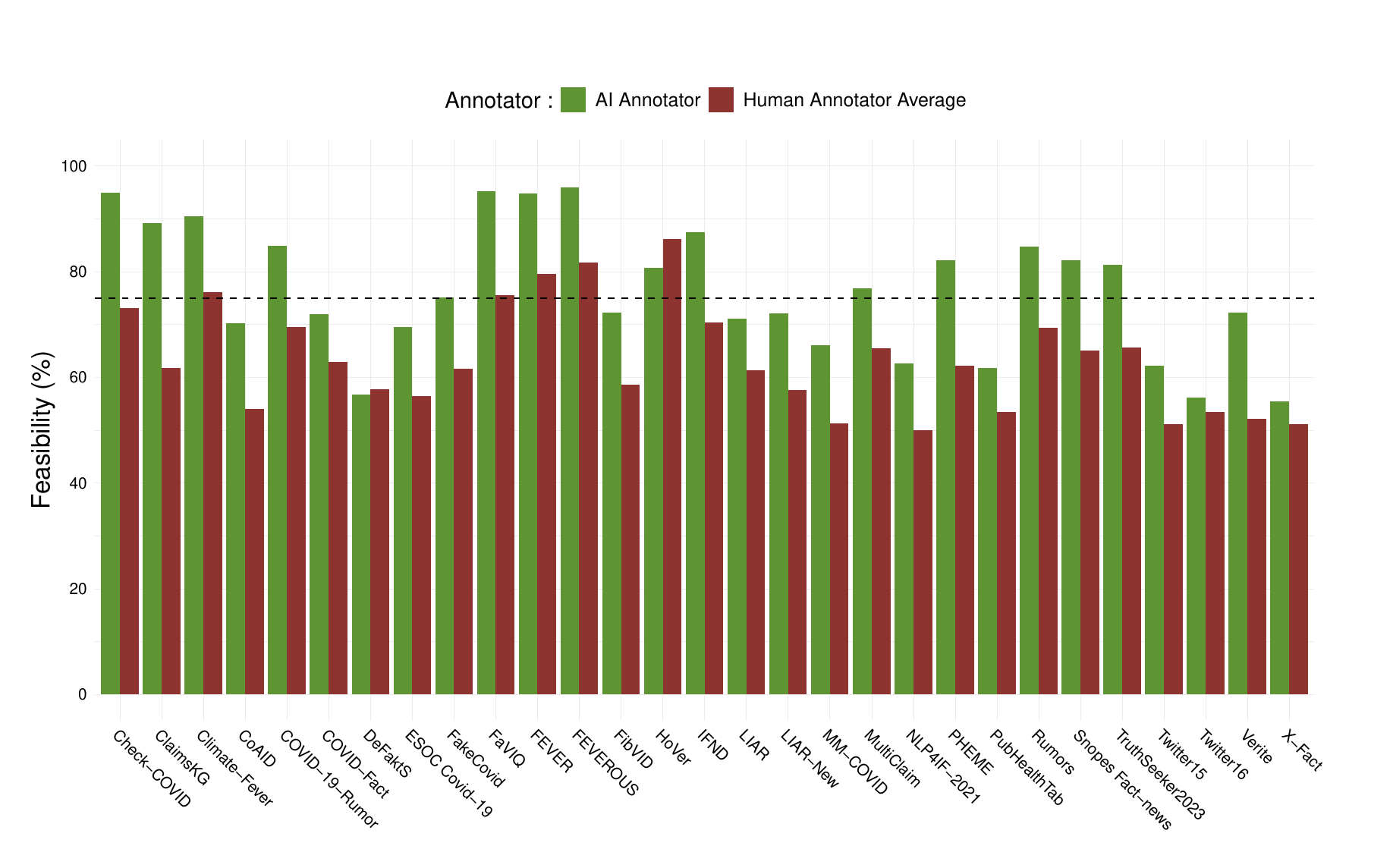}
        
    \vspace{-3mm}
        \caption{Evaluation of claims dataset feasibility. Even with evidence retrieval, most datasets have a concerning proportion of infeasible data.}
    \label{fig:feasibility_raw_data}
\end{figure*}

\begin{figure}[htbp]
    \centering
    \includegraphics[width=\columnwidth]{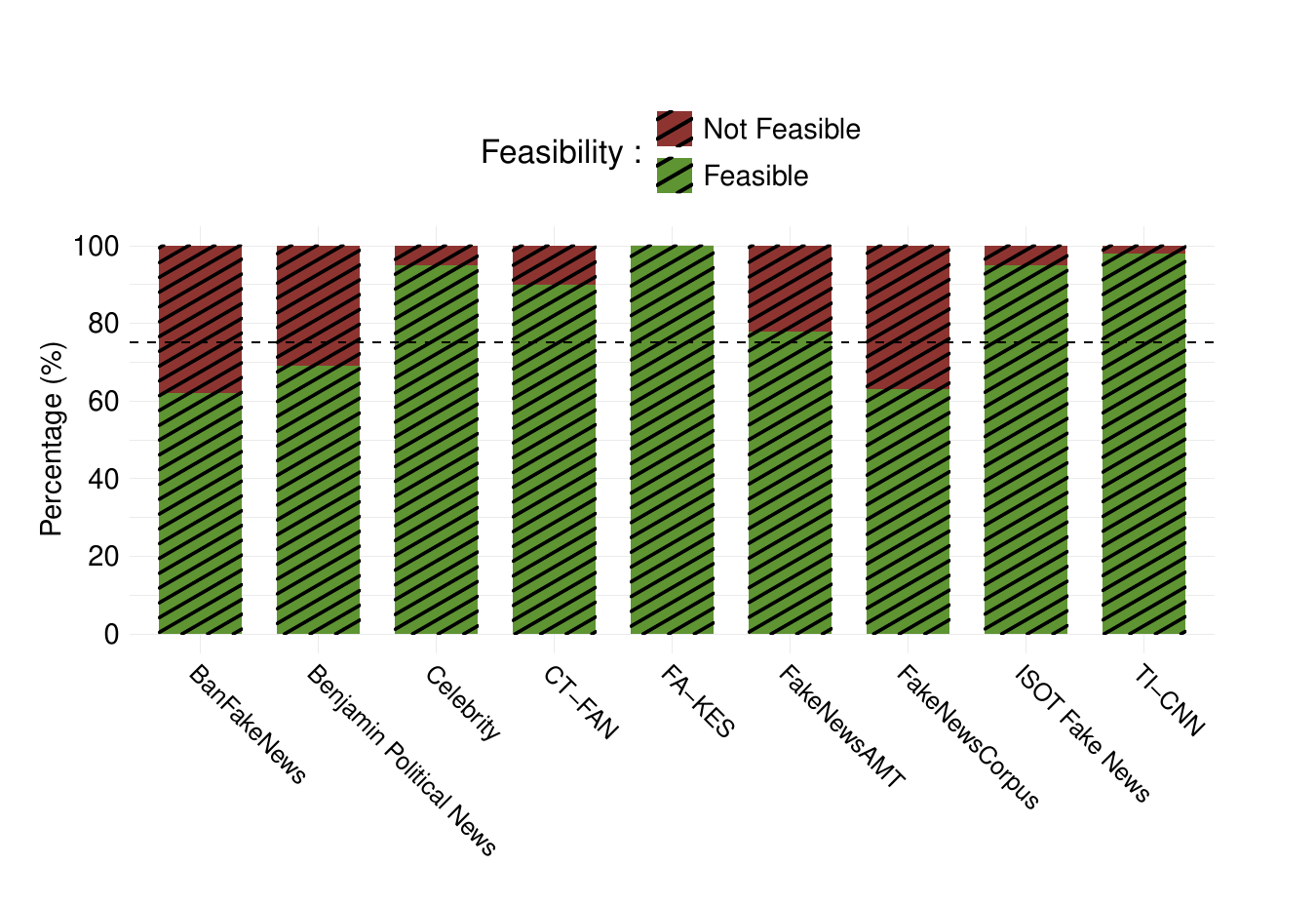}
    \caption{AI annotator evaluation of feasibility over paragraph datasets. These datasets often have higher feasibility proportions than claims datasets.}
    \label{fig:feasibility_paragraph}
\end{figure}

We next focus on systems that \emph{do} have access to retrieval, particularly open web search, and assess which datasets will have strong feasibility and therefore represent strong options for training and evaluation. We propose 75\% feasibility as a generous threshold, allowing for a moderate amount of noise in both our assessment and in applications relying on the datasets. We focus on the human annotator average (detailed in Appendix~\ref{app:feasibility}), noting that the AI annotator has decent alignment with human assessment, and advantages in sample size and scalability, but is typically a bit more generous. Regardless, both forms of assessment yield similar conclusions. We note that for some applications, up to a quarter of examples in the evaluation data representing noise may be far too much---for example, comparisons between methods with margins of a couple percentage points. Nonetheless, we see in Figure~\ref{fig:feasibility_raw_data} that most datasets do not even meet this standard.

We next turn to paragraph datasets. We find (Figure~\ref{fig:feasibility_paragraph}) that they have a higher feasibility rate than claims datasets. This result makes sense considering that having a greater volume of text allows for more context, which reduces the amount of ambiguity in classifying misinformation. Thus, while it can be more complex to assess veracity with longer inputs, evaluation in this setting may be paradoxically easier. Since recent advances in LLMs (and multimodal systems that can process other types of inputs) provide many options for working with longer data, prioritizing this and other context-rich settings may be beneficial for the field.

We also observe in Appendix~\ref{app:feasibilitypredictability} that feasibility has a substantial relationship (e.g., 0.56 Pearson correlation) with the performance of a veracity prediction system \emph{across} different datasets. We note a caveat that it appears very noisy \emph{within} datasets, i.e., feasibility is not a strong predictor for whether an individual example's veracity will be assessed correctly, This is likely because there are many reasons beyond feasibility that a prediction system can be correct or incorrect, such as data leakage, lucky search results or assumptions about context or predictions in general, etc.

Overall, there are 3 main implications. First, there is an urgent need for higher quality claims datasets. Second, supplementary information beyond claim text, such as claim dates, authors, extended paragraphs, or other additional context may help alleviate this problem. However, it is critical to assess how the specific information a predictive method processes impacts the feasibility of the data it is being trained and evaluated on, to confirm that performance margins between methods represent real progress and not progress in predicting noise. Thus, third, we propose that even in the more favorable retrieval-augmented setting, Evaluation Quality Assurance (Section~\ref{sec:eqa}) is paramount.

%% file: 050detection_eval.tex
\section{Evaluation}

\subsection{Baseline Performance}

\input{tables/search_baselines}

We next discuss baselines when using these datasets. Recent works have shown LLMs represent the state-of-the-art for misinformation detection on generic claims \cite{chen2023combating, pelrine2023towards}. But despite its importance \cite{pelrine2021surprising}, both human and compute time constraints can be a barrier to comparing with such strong baselines.

Thus, we provide two new baselines for future use. We follow the recent method of \cite{tian2024web} and use GPT-4-0125 in two ways: directly prompting the LLM for a veracity evaluation, and providing the LLM a web search tool to first collect evidence before forming a final verdict. We note that although these are state-of-the-art systems for zero-shot misinformation detection, they should not be regarded as sole or permanent points of comparison. Stronger LLMs and methods could replace them eventually. Nonetheless, they can provide a useful point of comparison for the near future. 

We note that 8 datasets are excluded from this baseline: 7 tweet datasets that we were unable to retrieve due to X API limits, and the ESOC Covid-19 dataset because it only has a ``refutes'' label. Results on all others are provided in Table~\ref{tab:baselines}. Notably, because these are zero-shot approaches, they are much less vulnerable to spurious correlations than models trained on each of these datasets, sometimes leading to dramatically lower but more realistic performance compared to alternatives in the literature (e.g., Twitter15 and Twitter16, where temporal classification achieves over 80\% F1).

\subsection{The Flaw in Current Metrics}
\label{sec:eval-flaw}

When looking at the outputs of the prediction system, we observe cases where the predicted label did not match the ground truth, yet the evidence and reasoning of the system was valid. For instance, in one example on the FEVER dataset, the input claim is ``Vietnam is a place'' and the prediction said roughly ``Vietnam is not just a place, it's a country!'' In another example from LIAR-New, a statement was marked false by PolitiFact because it was in the context of a fake video, but the statement itself did not mention the video and in isolation would be true. In cases like these (and further examples in Appendix~\ref{app:labelevalexamples}), a simple binary or categorical label cannot provide an informative evaluation.

To determine the prevalence of this phenomenon, two human expert annotators evaluated (Appendix~\ref{app:manuallabel}) chain-of-thought rationales from the web-search enabled baseline prediction system \cite{tian2024web}.

\input{tables/full_text_eval}

We observe a consistently high false-incorrect rate (first column of Table~\ref{tab:full_text_annotation}) and a generally low false-correct rate (fourth column of Table~\ref{tab:full_text_annotation}).
Therefore, when benchmarking generative AI misinformation detection systems using categorical labels, the predictive accuracy and similar metrics reflect a reasonable lower bound on the performance---but a terrible upper one, that marks many valid responses incorrect. 
We also observe that there is a large amount of ambiguity and room for interpretation in the examples that are being marked wrong by categorical label in these three datasets. 
Hence: 
\begin{enumerate}[leftmargin=10pt,topsep=2pt,noitemsep]
    \item Categorical metrics cannot be used alone to compare generative and non-generative systems. Although multiple recent works (e.g., \citet{pelrine2023towards, chen2023combating, wei2024long, yu2023open}) have highlighted the effectiveness of recent LLMs for misinformation detection, their comparisons with prior approaches may even still be underestimating the dominance of LLMs in this domain.
    \item Generative systems need many, repeated, and large-margin measurements if the categorical lower bound alone is to form meaningful comparisons between them.
    \item There is an urgent need for better datasets and better evaluation procedures in this domain that are suitable for the generative AI era. 
\end{enumerate} 

Although to our knowledge not addressed in the context of misinformation detection on claims, challenges in evaluations of generative AI have been broadly documented in other fields \citep{mcintosh2024inadequacies,ahuja2023mega, michel2023challenges,basole2024generative}. A common approach aimed at solving this is having an LLM assess the full prediction rationales \citep{bai2022constitutional,sun2023principle}, with \citet{sun2023salmon} noting that LLM-powered evaluation can produce more consistent preference signals than human annotators. In general, using LLMs for evaluation enables one to leverage much richer signals than simple categorical predictions and labels, while avoiding reliance on often inaccessible human evaluators.

As an initial step towards higher-fidelity evaluation, we constructed an evaluator based on contradictions between the explanation generated by a predictive system, and a fact-checking article. In particular, we provide GPT-4-0409 both \textit{the prediction} and \textit{the fact-checking article}, and ask it to score contradictions from 0 (no contradiction) to 10. The exact prompt and other implementation details are provided in Appendix~\ref{app:contradictioneval}. We chose a score-based approach to avoid forcing a potentially misleading binary in cases where there is a partial contradiction. With this approach, good predictions should have low contradiction against a high quality, professional fact-checking article. We also tested binary and trinary versions of this prompt, described in the Appendix, which yielded nearly identical results. 

To evaluate this evaluator, we set the oracle-optimal threshold of 3 or less (low contradiction) to indicate a prediction that the rationale is not wrong, and 4 or more indicating one that the rationale is wrong. We find that the contradiction score evaluator agrees 68\% of the time with the human labels of valid and invalid rationales (reported in Table \ref{tab:full_text_annotation}) on the LIAR-New dataset. This is higher than the 60\% original human inter-annotator agreement on these rationales (before disagreement resolution described in Appendix~\ref{app:manuallabel}), and suggests the method extracts a meaningful but not definitive evaluation signal. We also note, though, that there is more to high quality misinformation detection than just a lack of contradiction. Therefore, we do not suggest using this tool as a primary evaluator. But we provide these results, along with all the data---inputs, predictions, manual labels, fact-checking articles, and evaluator outputs---on our GitHub, as a potential springboard for stronger evaluation methods in future research.

%% file: tables/search_baselines.tex
\begin{table}[]

    \caption{State-of-the-art GPT-4 baselines, with and without web search.}
    \centering
    \begin{tabular}{lcc}
    \toprule
    Dataset & F1 (Search) & F1 (Offline) \\
    \midrule
    Check-COVID &  78.8\% &  85.4\%  \\
    Climate-Fever &  66.9\% &  65.3\%  \\
    CoAID &  62.0\% &  60.3\%  \\
    COVID-19-Rumor &  62.8\% &  65.8\%  \\
    COVID-Fact &  67.5\% &  67.4\%  \\
    FakeCovid &  50.4\% &  51.0\%  \\
    FaVIQ/test &  81.5\% &  80.7\%  \\
    FEVER/paper\_test & 88.6\% & 89.2\% \\
    FEVEROUS/validation & 65.6\% & 62.2\% \\
    FibVID &  67.6\% &  67.3\%  \\
    HoVer/validation & 68.8\% & 61.7\% \\
    IFND &  56.5\% &  42.0\%  \\
    LIAR/test &  44.8\% &  50.7\%  \\
    LIAR-New & 69.7\% & 63.6\% \\
    MM-COVID &  85.6\% &  86.5\%  \\
    PHEME &  34.3\% &  33.4\%  \\
    PubHealthTab/test &  30.8\% &  49.3\%  \\
    Rumors &  69.5\% &  80.7\%  \\
    Snopes Fact-news &  90.6\% &  81.4\%  \\
    TruthSeeker2023 &  81.9\% &  81.0\%  \\
    Twitter15 &  57.7\% &  66.5\%  \\
    Twitter16 &  49.2\% &  55.8\%  \\
    Verite &  63.3\% &  59.8\%  \\
    X-Fact/test &  55.0\% &  53.0\%  \\
\bottomrule
\end{tabular}
    \label{tab:baselines}
    \vspace{-1mm}
\end{table}

%% file: tables/full_text_eval.tex

\begin{table}[h]
\centering
\caption{Agreed-upon manual annotations. Many predictions marked invalid by simple comparison with the categorical labels are actually valid. Standard evaluations are likely to systematically under-evaluate the performance of generative systems.}
\label{tab:full_text_annotation}
    \begin{tabular}{l c c c c}
    \toprule
{}~  & \multicolumn{2}{c}{\ensuremath{\text{Label} \ne \text{Prediction}}} & \multicolumn{2}{c}{\ensuremath{\text{Label} = \text{Prediction}}} \\
    \cmidrule(lr){2-3} \cmidrule(lr){4-5}
 AI rationale is:     & Valid & Invalid & Valid & Invalid \\
    \midrule
    LIAR-New & 55/100 & 30/100 & 76/100 & 1/100  \\
    FEVER  & 38/100 & 34/100 & --    & --   \\
    MM-COVID   & 39/70  & 3/70   & 89/100 & 0/100  \\
    \bottomrule
    \end{tabular}

\end{table}

%% file: 060conclusion.tex
\section{Evaluation Quality Assurance}
\label{sec:eqa}

We have highlighted multiple ways datasets and evaluation procedures could deliver results that will fail to translate to the real world. While the information here directly reveals some pitfalls, we suspect that future research---with new datasets, new methods, and new needs---will encounter similar problems. To improve the validity of conclusions in our field in a systematic and future-proof way, we propose \emph{Evaluation Quality Assurance (EQA)} as a key component of an improved research methodology. 

We draw inspiration from the common practices for systematically and transparently documenting dataset publications, such as ``datasheets for datasets''~\cite{gebru2021datasheets}, and suggest that similar attention should be given to the \emph{usage} of datasets. However, it is important to ensure data and evaluation quality not just as a lengthy post-hoc appendix, but as an integral part of the research process and a key foundation of every paper's core arguments. Therefore, we propose including EQA as a scientifically essential step within the methodology. This means augmenting the standard \emph{descriptive} discussion of data used with a \emph{critical} analysis of it, explaining what steps and quality assessments were conducted to assure that the data is suitable to prove the experimental conclusions. We suggest that dedicating even just one paragraph to this, rather than taking quality of ``standard'' datasets for granted, could dramatically improve the validity of our empirical results and the rate of real progress in the field. Conversely, current practices risk misleading evaluations, generalization failures, and immeasurable time and effort dedicated to measurements of noise.

We make two key recommendations for Evaluation Quality Assurance. 

\begin{itemize}[left=10pt,topsep=2pt,noitemsep]

\item \textbf{Quality Assessment}: analyze the data and evaluation processes themselves, not just their results, to assure they are suitable to test performance and generalization of a method. One basic analysis, that can be nearly universally recommended, is to qualitatively inspect at least 25 random examples---input, output, and ground truth---where the method made correct predictions, and 25 where it made incorrect ones. This data can be shared for reproducibility, and provides a basic sanity check. This simple step could already have a high impact; for example, the feasibility and categorical metrics issues we have reported would be detected by such a sanity check in the majority of the datasets that we examined. Furthermore, we also recommend papers perform quantitive assessments of their data and evaluation, such as (but not limited to) the keyword, temporal, and feasibility assessments discussed in prior sections and available in our \href{https://github.com/ComplexData-MILA/misinfo-datasets}{\textbf{\cdlme} toolkit}.

\item \textbf{Assurance Limitations}: report \textbf{limitations in evaluation quality assurance (EQA)}, not only in terms of dataset size or experimental scope but also in terms of quality issues that the EQA could not rule out. Key to this is considering how a method would be applied in the real world---and how the evaluation might not be fully capturing the real-world conditions. If, for instance, a method is intended to handle diverse contexts or domains of misinformation, but was only tested on data from a single context (e.g., one country) or domain (e.g., politics), this should be stated as a limitation so that future research can address it. Explicitly acknowledging such gaps strengthens scientific rigor and ensures that claims are appropriately supported by the evidence provided.


\end{itemize}

In this way, we treat assuring evaluation quality as a crucial step for validating the research process. For authors, this will help reinforce the demonstrated rigour of the experiments and the validity of proposed methods. It will also prevent the rejection of strong methods due to unrealistically high baseline settings caused by spurious correlations or other evaluation issues. For reviewers, this will produce more grounding (or expose the lack thereof) for the experiments, as well as a framework for recommendations when evaluation is insufficiently justified and needs better validation. Thus, we propose that adopting EQA can efficiently and substantively mitigate propagating and compounding errors, that arise in this domain from current practices of copy-pasting evaluation from past literature.

\section{Related Work}

In recent years, the scientific community has devoted considerable research to curbing misinformation, defined broadly as ``false or misleading information" \citep{lazer2018science}. Initial surveys, such as \citet{shu2017fake}, provided foundational frameworks and overviews of detection techniques from data mining perspectives, while subsequent studies by \citet{oshikawa2018survey} and \citet{zhou2020survey} emphasized NLP methods and interdisciplinary approaches. Despite these contributions, most existing surveys \citep{bondielli2019survey, gravanis2019behind} largely overlook detailed analyses of misinformation datasets, offering superficial coverage that neglects dataset-specific qualities essential for accurate model development and validation. Recent surveys partially address this gap by categorizing datasets quantitatively \citep{d2021fake, sharma2019combating, ali2022fake, patra2022understanding}, yet typically lack in-depth analysis. Other works, including \citet{abdali2024multi}, \citet{pelrine2021surprising}, and \citet{wu2022probing}, provide deeper insights into dataset biases and limitations but remain limited by analyzing only a small subset of datasets. To address these limitations, our work evaluated a broad selection of misinformation datasets, assessed in depth how issues in data and evaluation could lead to misleading conclusions, and provided tools and recommendations for future improvements in the field.

\section{Limitations}

First, we note that while to our knowledge this represents the largest and most comprehensive survey of datasets in this domain, there are certainly many other datasets in existence and it is probable that some were not included. There is also a steady stream of new datasets being created every year. In the near future, we plan to collect external feedback and update our survey to maintain and expand the comprehensiveness of our study.

We also note that our unified label schema simplifies some labels that might have meaningful information, for example, gradations of veracity instead of binary true/false. Some granularity has been traded for the ability to establish a unified schema across all the claims datasets. When using these datasets, we advise careful consideration of the optimal labels to apply.

As discussed previously, additional work is needed in evaluation, both to confirm that the observed validity issues with metrics like accuracy are widespread (as we hypothesize) and to create strong, thoroughly tested alternatives. We also note that the baselines we have provided use old evaluation procedures on LLM-based predictors. This can be flawed both for the reasons discussed in Section~\ref{sec:eval-flaw}, and potentially also because a substantial proportion of the data could be within the LLM training data. \citet{pelrine2023towards} indicates LLM-based methods offer the strongest performance even beyond their knowledge cutoffs, and using web search to actually provide evidence can mitigate this to some degree. But nonetheless, these baselines should be viewed carefully and with due attention to both their strengths and limitations, and future work to establish more universal baselines---as well as datasets and evaluation methods that enable them---would be very valuable.

Lastly, although we discussed multiple key dimensions of misinformation detection datasets, a favorable assessment in these dimensions does not guarantee a dataset suitable for any application. Other types of limitations in the data could still lead to spurious shortcuts instead of generalizable predictions and evaluation. Our proposed EQA procedure can help detect this with respect to particular methods and applications, but we also encourage future work to identify additional limiting factors in evaluation in this domain, and to solve them.

\section{Conclusion}

High quality data and evaluation are essential for realistic results and rapid progress in the field. In this work, we have provided a guide to misinformation detection datasets aiming at both quantity and quality. We also highlighted limitations of existing datasets and evaluation approaches, which may have spurious correlations, infeasible examples, and misleading results. We hope that this work can provide a roadmap to better grounding for future predictive methods research that needs to select datasets and evaluation approaches. Meanwhile, we also hope that this guide will provide foundational understanding and a call to action to build new and better datasets and evaluation procedures.

\section*{Acknowledgements}
This project has been partially funded through the Canada CIFAR AI Chair program and the Centre for the Study of Democratic Citizenship. Kellin Pelrine was supported by funding from the Fonds de recherche du Queb\'ec.

\section*{Author Contributions}

Camille Thibault led the survey and curation of all the datasets. Jacob-Junqi Tian led all LLM-based experiments. Both first authors contributed across diverse other areas of the project. Gabrielle Peloquin-Skulski made key contributions to dataset curation and unification, and the early direction of research. Taylor Lynn Curtis led the public website and repositories for the data from prototypes to completion, and contributed to writing. James Zhou led paragraph datasets. Florence Laflamme contributed human analysis and annotation. Luke Yuxiang Guan contributed to the website and data repository. Reihaneh Rabbany and Jean-François Godbout advised the project. Kellin Pelrine advised and managed the project. Pelrine also proposed foundational ideas, including each quality assessment, the flaw in current metrics, EQA, and the research direction overall.

%% file: 070appendix.tex
\newpage

\input{tables/71_data_description}

\input{030survey}

\subsection{Dataset Details}

This section provides an overview of all claims datasets and paragraph datasets. The number of entries, the collection method, and the original labels are discussed.

\subsubsection{Claims Dataset Details}
\label{app:datadetails}

\textbf{AntiVax} \citep{hayawi2022anti}: AntiVax is a dataset containing 15,465,687 tweets about the COVID-19 vaccine, of which only 15,073 are annotated for the model training. These were collected via the Twitter API form December 1, 2020 to July 31, 2021. The annotations are binary (misinformation or not misinformation). Tweets labeled as misinformation include opinions or general news about the vaccine. Tweets containing sarcasm or humor are not classified as misinformation.

\textbf{Check-COVID} \citep{wang2023check}: This dataset contains 1,504 expert-annotated claims about the COVID-19 pandemic. These claims are either composed by annotators or extracted from news articles. Each claim is also paired with a sentence evidence from scientific journals. Labels are divided into three categories: support, refute or not enough info.

\textbf{ClaimsKG} \citep{tchechmedjiev2019claimskg}: ClaimsKG is a dataset of 74,066 claims published between 1996 and 2023. These claims were collected from 13 fact-checking sites and annotated as true, false, mixture or other. It is essential to point out that the version of ClaimsKG we used for the analysis contains 67,009 claims. It was provided by the authors.

\textbf{Climate-Fever} \citep{diggelmann2020climate}: Climate-Fever is a dataset about climate change. It includes 7,675 annotated claim-evidence pairs. Claims are collected on the Internet while evidences are retrieved from Wikipedia. Each claim is assigned one of the following labels: supports, refutes or disputed.

\textbf{CMU-MisCOV19} \citep{memon2020characterizing}: CMU-MisCov19 is a dataset about COVID-19. It contains tweets that were collected over three days: March 28, 2020, June 15, 2020, and June 24, 2020. 4,573 tweets are annotated based on various types of information and misinformation. In total, there are 17 categories, such as irrelevant, conspiracy, true treatment, fake cure, false fact, ambiguous, etc.

\textbf{CoAID} \citep{cui2020coaid}: This dataset covers various COVID-19 healthcare misinformation. It contains 4,251 news, 926 social platforms posts, and 296,000 related user engagements. All facts are collected between December 1, 2019 and September 1, 2020. All the data is annotated in a binary form: true or fake.

\textbf{Counter-covid-19-misinformation} \citep{micallef2020role}: Covering four-month period, this dataset contains 155,468 tweets relating to COVID-19 and, more specifically, fake cures and 5G conspiracy. The tweets were harvested from an existing dataset \footnote{\url{https://doi.org/10.2196/19273}} and Twitter. 4,800 claims are annotated, and the labels are divided into three categories: misinformation, counter-misinformation, or irrelevant.

\textbf{COVID-19-Rumor} \citep{cheng2021covid}: This dataset includes 7,179 annotated claims crawled from Google and Twitter from January 2020 to March 2020. The topics of these claims, all related to COVID-19, include emergency events, comments from public figures, updates on the coronavirus outbreak, etc. The labels were manually assigned and cross-validated. The labels are also divided into three categories, consisting of true, false, or unverified.

\textbf{Covid-19-disinformation} \citep{alam2020fighting}:  This is another dataset about COVID-19 disinformation. It contains 16K coded claims in Arabic, Bulgarian, Dutch, and English. These were collected via the Twitter API between January 2020 and March 2021. Their labels are fined-grained. The annotation task involved determining the truthfulness of the tweet, its potential to cause harm, whether it is relevant for policymakers, etc.

\textbf{COVID-Fact} \citep{saakyan2021covid}: Also on the subject of COVID-19, Covid-Fact contains 4,086 claims. Among these, 1,296 are factual claim from the $r/COVID19$ subreddit, while 2,790 are false claims automatically generated. All claim contain evidence, and the labels are binary: supported or refuted.

\textbf{Covid-vaccine-misinfo-MIC} \citep{kim2023covid}: Covid-vaccine-misinfo-MIC is a geolocated and multilingual dataset about COVID-19. It spans from 2020 to 2022, and includes 5,952 tweets from Brazil, Indonesia, and Nigeria. The claims are all labeled in a granular form, indicating whether they are vaccine-related, contain misinformation, are political, etc.

\textbf{DeFaktS} \citep{ashraf2024defakts} : DeFaktS is a database of 105,855 claims from X (formerly Twitter), of which 20,008 are annotated. Claim topics are varied. They include, for example, war in Ukraine, elections, covid-19 pandemic, energy crisis, climate, inflation, etc. All the claims are written in German and the veracity labels are fine-grained, as they include binary labels (real, fake) and labels stating content, authenticity, psychology and semantic features.

\textbf{ESOC Covid-19} \citep{siwakoti2021localized}: ESOC contains 5,613 claim-stories about misinformation gathered from the early days of the COVID-19 pandemic up to the end of December 2020. These claims come from all five continents and all contain misinformation.

\textbf{FakeCovid} \citep{shahi2020fakecovid}: FakeCovid is a dataset containing news claims about COVID-19. These data were collected from 92 different fact-cheking websites between January 4, 2020, and May 15, 2020, covering 40 languages and originating from 105 countries. The truthfulness labels (false, mostly false, misleading, half true, mostly true, no evidence) are derived from experts at fact-checking agencies. The dataset also includes other labels defining the type of false news (prevention \& treatments, international response, conspiracy theories, etc), all annotated by members of their team.

\textbf{FaVIQ} \citep{park2021faviq}: This dataset contains 188K annotated claims and evidences. Each claim has been converted based on questions from the Google Search queries. The claims cover various subjects including culture, sports, and history. The labels are binary: support or refute.

\textbf{FEVER} \citep{thorne2018fever}: This dataset includes 185,445 coded claims generated by altering sentences extracted from the 50,000 most popular Wikipedia pages. Annotators were tasked with crafting claims covering a wide array of topics, ranging from historical facts to entertainment trivia, each containing a single fact. The labels assigned to these claims were determined based on evidence sourced from Wikipedia as well, and they were categorized in a binary manner as either supported or refuted.

\textbf{FEVEROUS} \citep{rami2021fact}: Continuing in the same vein as FEVER, FEVEROUS is a dataset containing 87,026 claims extracted from Wikipedia. Each claim is annotated based on associated evidence. One distinctive feature with FEVER is that the labels are divided into three categories: supported, refuted, or not enough information.

\textbf{FibVID} \citep{kim2021fibvid}: This COVID-19 related dataset was collected by crawling 1,353 news claims and the labels of two fact-checking websites, Politifact and Snopes. These news claims were subsequently matched with 221,253 relevant tweets written by 144,741 users between February 1, 2020 and December 31, 2020. The labels from the fact-checking websites were simplified in a binary manner, classifying them as either true or false.

\textbf{HoVer} \citep{jiang2020hover}: This dataset contains 26,171 claims covering various topics. These claims are derived from question-answer pairs sourced from the HOTPOTQA dataset \footnote{https://doi.org/10.18653/v1/D18-1259}. Annotators from Appen3 were trained to rewrite these question-answer pairs to a single sentence. To determine the veracity labels, the authors extracted facts from Wikipedia and asked the same annotators to label the claims based on whether they supported them or not.

\textbf{IFND} \citep{sharma2023ifnd}: The Indian Fake News Dataset (IFND) consists of texts and images collected between 2013 and 2021. These data cover elections, politics, COVID-19, violence, and miscellaneous topics. The veracity of these data is determined based on the media from which they were collected. True claims originate from Tribune, Times Now News, The Statesman, and others, while false claims come from the fact-checked columns of Alt News, Boomlive, and media outlets like The Logical Indian, and News Mobile.

\textbf{LIAR} \citep{wang2017liar}: LIAR is a dataset of 12.8K short statements scraped from the API of Politifact, a fact-checking website. These statements were made by politicians and can cover various subjects including the economy, health care, and the job market. All of these political statements were manually labeled by Politifact journalists. The truthfulness ratings consist of six categories: pants-fire, false, barely true, half-true, mostly true, and true.

\textbf{LIAR-New} \citep{pelrine2023towards}: Liar-New is a dataset containing 1,957 claims scraped from Politifact over a period dating from October 2021 to November 2022. Like Liar, these statements focus on the American political class and encompass various topics including health, the economy, and education. Each claim has also been translated into French by two native speakers. Veracity labels are issued by Politifact's fact-checkers and consist of 6 categories: pants-fire, false, barely true, half-true, mostly true, and true. Unlike Liar, Liar-New features possibility labels (possible, impossible or hard). These labels identify whether claims have enough context to be verified.

\textbf{MediaEval} \citep{boididou2015verifying}: This dataset was made available for the MediaEval 2015 test. It includes tweets and images concerning 11 events, such as Hurricane Sandy, the Boston Marathon bombing, the Sochi Olympics, and the Malaysia Airlines Flight 370. The labeling approach is binary. A tweet is labeled as real if it shares multimedia that accurately represents the referenced event, whereas a tweet is labeled as fake if it shares multimedia content that misrepresents the referenced event.

\textbf{MM-COVID} \citep{li2020mm}: MM-COVID is a dataset containing claims from 6 languages: English, Spanish, Portuguese, Hindi, French, and Italian. The data and their labels were crawled from fact-cheking agencies and reliable media sources. Each claim was then matched with social media engagements from Twitter users. The labels are binary (real or fake).

\textbf{MultiClaim} \citep{pikuliak2023multilingual} : Multiclaim contains 31,305 claims from social media posts in 39 languages. Each of these claims is associated with an article and a label issued by a fact-checking website. The subjects are diverse, and the database also includes a translation of all claims into English.

\textbf{NLP4IF-2021} \citep{shaar2021findings} : NLP4IF-2021 is a database of 3,172 Covid-19 X claims. Three languages are present in NLP4IF-2021: Arabic, Bulgarian and English. The veracity labels are binary (yes or no to the question \textit{To what extent does the tweet appear to contain false information?}) and the dataset also contains other labels covering, for example, its harmfulness, its interest for the general public and its need to be fact-checked by experts.

\textbf{PHEME} \citep{Zubiaga2016}: This dataset contains tweets published during five breaking news periods: Charlie Hebdo, Ferguson, Germanwings Crash, Ottawa Shooting, and Sydney Siege. Each tweet is annotated as either a rumor or non-rumor.

\textbf{PubHealthTab} \citep{akhtar2022pubhealthtab}: This dataset contains 1,942 real-world claims about public health. These claims are extracted from fact-checking and news review websites. Each claim is associated with a summary of the article, a veracity label, and a justification for that label. The labels are coded into three categories: support, refute or not enough info.

\textbf{Rumors} \citep{tam2019anomaly}: Rumors is a dataset containing 1,022 rumors collected between May 1, 2017, and November 1, 2017 from the fact-checking website Snopes. The rumors cover various topics, including politics, fraud, fauxtography, crime, and science. Each claim is also associated with tweets, and the veracity labels are as follows: true, mostly true, mixture, mostly false, false, unproven.

\textbf{Snopes Fact-news} \citep{snopes_factnews_data}: This dataset is scraped from the fact-checking website Snopes. It contains 4,550 claims, all associated with veracity labels, the origin of the claim, a summary of this origin, and short descriptions of what is true and what is false. The labels are the same as RUMORS, namely true, mostly true, mixture, mostly false, false, unproven.

\textbf{TruthSeeker2023} \citep{dadkhah2023largest}: TruthSeeker2023 is a dataset of 180,000 coded claims from 2009 to 2022. To collect them, the authors initially crawled 1,400 claims and their ground-truth labels from Politifact. Then, keywords from these claims were used to collect associated tweets, which crowdworkers verified for accuracy. These tweets were labeled based on their corresponding claims from Politifact. TruthSeeker2023 includes two label types: a five-way label (Unknown, Mostly True, True, False, Mostly False) and a three-way label (Unknown, True, False).

\textbf{Twitter15} \citep{ma2017detect} : Twitter15 contains 1,490 tweets. To identify fake news, two rumor tracking websites, Snopes and Emergent, were used. Tweets related to these fake news stories were then scraped from Twitter using keywords, and their matches were cross-checked by three researchers. Real news tweets was also collected from Twitter via Twitter's free data stream. It's important to note that this is not the original dataset. The original \citep{liu2015real} has been re-used by the authors of this new database, who have kept the same name while modifying only the labels. The veracity labels are “true”, “false” and “non-rumor”. To classify them, \citet{ma2017detect} has labeled them according to whether or not the author denies the rumor.

\textbf{Twitter16} \citep{ma2017detect}: Twitter16 is a dataset containing 818 tweets. Like Twitter15, Twitter16 was reproduced by \citet{ma2017detect}. For the original dataset \citep{ma2016detecting}, the authors followed the same data collection procedure as for the original Twitter15, but focused solely on the collection of fake news using Snopes. \citet{ma2017detect} have modified the labels, which are true, false, unverified, and non-rumor.

\textbf{Verite} \citep{papadopoulos2024verite}: VERITE is a dataset containing 1,001 claims and associated images. The data were collected from Snopes and Reuters from January 2001 to January 2023. The topics covered are diverse, including politics, culture, entertainment, business, sports, environment, religion, and more. The labels, derived from fact-checking agencies, are coded into three categories: true, out-of-context, and miscaptioned.

\textbf{WICO} \citep{pogorelov2021wico}: WICO is a dataset dedicated to COVID-19. It includes 364,325 claims. These claims were collected via the Twitter API from January 17, 2020, to June 30, 2021. Approximately 10,000 tweets are manually annotated with the following labels: 5G conspiracy, other conspiracy, non-conspiracy, and undecidable.

\textbf{X-Fact} \citep{gupta2021x}: X-FACT is a dataset of 31,189 short statements scraped from 85 fact-checking websites. Covering various topics, the data are available in 25 languages, including Arabic, Bengali, French, Hindi, Indonesian, Italian, Spanish, Polish, and Portuguese. The veracity labels indicate a decreasing level of truthfulness: true, mostly true, partly true, mostly false, false, unverifiable, and other.

\subsubsection{Paragraph Dataset Details}
\label{app:paragraphdetails}

\textbf{BanFakeNews.} \citep{hossain2020banfakenewsdatasetdetectingfake}: BanFakeNews is a dataset containing approximately 50,000 news articles. The articles are in the Bangla language and are collected from 22 popular news portals in Bangladesh. The articles were labelled by crowd-sourcing. This dataset is annotated with a set of labels consisting of fake and authentic.

\textbf{BenjaminPoliticalNews.} \cite{horne2017justinfakenews}: BenjaminPoliticalNews is a dataset containing 296 news articles. These articles were collected from existing datasets and studies. The articles were labelled by a source-based method. This dataset is annotated with a set of labels consisting of real, satire, fake, and true.

\textbf{Celebrity.} \citep{pérezrosas2017automaticdetectionfakenews}: Celebrity is a dataset containing 500 news articles from magazines about celebrity gossip and hoaxes. The articles were labelled by a source-based method. This dataset is annotated with a set of labels consisting of legit and fake.

\textbf{CT-FAN.} \citep{shahi2021overview}: CT-FAN is a dataset containing 2462 news articles. These articles were collected from multiple fact-checking sites with news articles from 2010 to 2022. The articles were labelled by a source-based method. This dataset is annotated with a set of labels consisting of partially false, false, other, and true.

\textbf{FA-KES.} \citep{abu_salem_2019_2607278}: FA-KES is a dataset containing 804 news articles. These articles were collected from multiple fact-checking websites and social media platforms, encompassing multiple languages and domains. The articles were labelled by crowd-sourcing. This dataset is annotated with a set of labels consisting of true, authentic, and fake.

\textbf{FakeNewsCorpus.} \citep{pathak-srihari-2019-breaking}: FakeNewsCorpus is a dataset containing 9408908 articles. These articles were collected from a curated list of 1001 domains. The articles were labelled by a source-based method. This dataset is annotated with a set of labels consisting of unreliable, fake, clickbait, conspiracy, reliable, bias, hate, junksci, political, unknown, and nan.

\textbf{FakeNewsAMT.} \citep{pérezrosas2017automaticdetectionfakenews}: fakenewsamt is a dataset containing 480 news articles. Real news articles were sourced from reputable news websites, while fake news articles were generated using Amazon Mechanical Turk, where crowdworkers were tasked with writing fictitious news content on given topics. This dataset is annotated with a set of labels consisting of fake and true.

\textbf{ISOT Fake News Dataset.} \citep{uvic2022fakenews}: ISOT Fake News Dataset is a dataset containing 44898 news articles. Real news articles were collected by crawling Reuters.com, while fake news articles were collected from unreliable websites identified by PolitiFact.com. This dataset is annotated with a set of labels consisting of true and false.

\textbf{TI-CNN.} \citep{yang2023ticnnconvolutionalneuralnetworks}: TI-CNN is a dataset containing 20015 news articles. These articles were collected from social media platforms and news websites. The articles were labelled by a source-based method. This dataset is annotated with a set of labels consisting of fake and real.

\subsection{Supplement on Labeling Approach}
\label{app:labelingapproach}

\input{tables/label_quality}

The task of annotating statements is both crucial and challenging for anyone attempting to train a robust classifier for misinformation detection. Precise labeling is essential to ensure the classifier's effectiveness, as it directly impacts its performance and reliability. Numerous approaches have been proposed in the literature to label true and false information. These approaches include expert and crowd-sourced annotation, source-based techniques, algorithmic methods, and a hybrid of these different approaches, all of which have been used in at least one of our 36 datasets (see Table~\ref{tab:factors}). We thus describe these different approaches used by the authors of the original datasets in turn to highlight their potential advantages and limitations.

\paragraph{Expert-based approach}

Experts and fact-checkers are a small group of non-partisan professionals from various disciplines who manually verify the veracity of information. The result of these verifications are often published in fact-checking websites such as \textit{Politifact} or \textit{Snopes}. The strength of this approach lies in its rigorous review process, ensuring each piece of information is thoroughly evaluated, which leads to consistent reviews across fact-checkers. However, this method is not scalable and is costly \citep{zhou2020survey}. As a result, experts must selectively choose the information they evaluate, which leads to many pieces of information going unchecked and potential biases in the selection of news and information that is evaluated \citep{lee2023fact, markowitz2023cross, walker2019republicans}. 

\paragraph{Crowd-sourced approach} 

Crowdsourced fact-checking involves enlisting non-expert laypeople to assess the accuracy of online information. These evaluations are then aggregated to determine the veracity of the content. This approach is advantageous because it is more scalable, and laypeople can respond to misinformation much more quickly than professional fact-checkers \citep{zhao2023variety}. Additionally, this method has been shown to be effective in reducing the spread of misinformation and to produce veracity ratings similar to those of professional fact-checkers \citep{allen2021scaling, martel2024crowds}. However, crowdsourcing also has its limitations. It can be challenging to filter out evaluations from non-credible users and to ensure a balanced representation of users from different partisan backgrounds \citep{zhou2020survey, martel2024crowds}.

\paragraph{Source-based approach}
Source-based approaches to verifying information involve evaluating the domain or author of the content.  Information is then rated as accurate if it comes from reliable sources and inaccurate otherwise. This method is more scalable than manual fact-checking, as it consists of evaluating the credibility of the source rather than each individual story. Additionally, this method is proven to be reliable, as experts generally rate news domains similarly \citep{lin2023high}. However, there are notable drawbacks. For instance, individual stories can vary in accuracy even within the same source, and not all content from low-quality outlets is necessarily false or misleading. Additionally, source familiarity significantly influences the perceived trustworthiness of content. Sources that are unfamiliar are often less trusted, which can lead to unfair negative evaluations of high-quality but lesser-known sources \citep{pennycook2019fighting, williams2023trustworthiness} 

\paragraph{Algorithmic methods} Finally, algorithmic methods can also be used to evaluate the veracity of content using NLP or other ML techniques \citep{zhou2020survey}. For example, Covid-fact uses a BERT-based classifier, FaVIQ uses T5-3B, and Rumors uses an approach based on a social graph. These methods offer significant advantages in scalability, as they can process vast amounts of data quickly and efficiently, making them suitable for large-scale verification tasks. However, their accuracy can be questionable in many cases, ranging from struggles with nuanced or context-specific content \citep{boukouvalas2024role}, issues with transfer and generalization \citep{huang2020conquering,pelrine2021surprising, pelrine2023towards}, or just generically poor performance (e.g., even state-of-the-art methods often have below 70\% accuracy compared to human labels \citep{zhang2023towards, pelrine2023towards}). Thus, the quality of algorithmic labels is often dubious.

\section{Data Quality Supplement}

\subsection{Supplement on Keyword Analysis}
\label{app:keywords}

Table \ref{tab:keywords_correlations} displays the results of the random forest classifier, including \textit{True and False} (T-F) labels, as well as \textit{True, False and Mixed} (T-F-M) labels for each dataset, alongside their corresponding baselines. Datasets located above the solid line represent the claim datasets and those positioned below the line correspond to the paragraph datasets.

\input{tables/keywords_correlations}

In Table~\ref{tab:keywords}, we show some examples of keywords that could lead to bad classifications. The number under the keywords is the number of times the word appears in claims based on its labels of veracity. We can thus see that there is an absence of true statements referring to Harris or Biden, but many that refer to Trump in the Truthseeker2023 dataset.

\begin{figure*}[h!]
    \centering
    \begin{minipage}{0.5\linewidth}
        \centering
        \includegraphics[width=\linewidth]{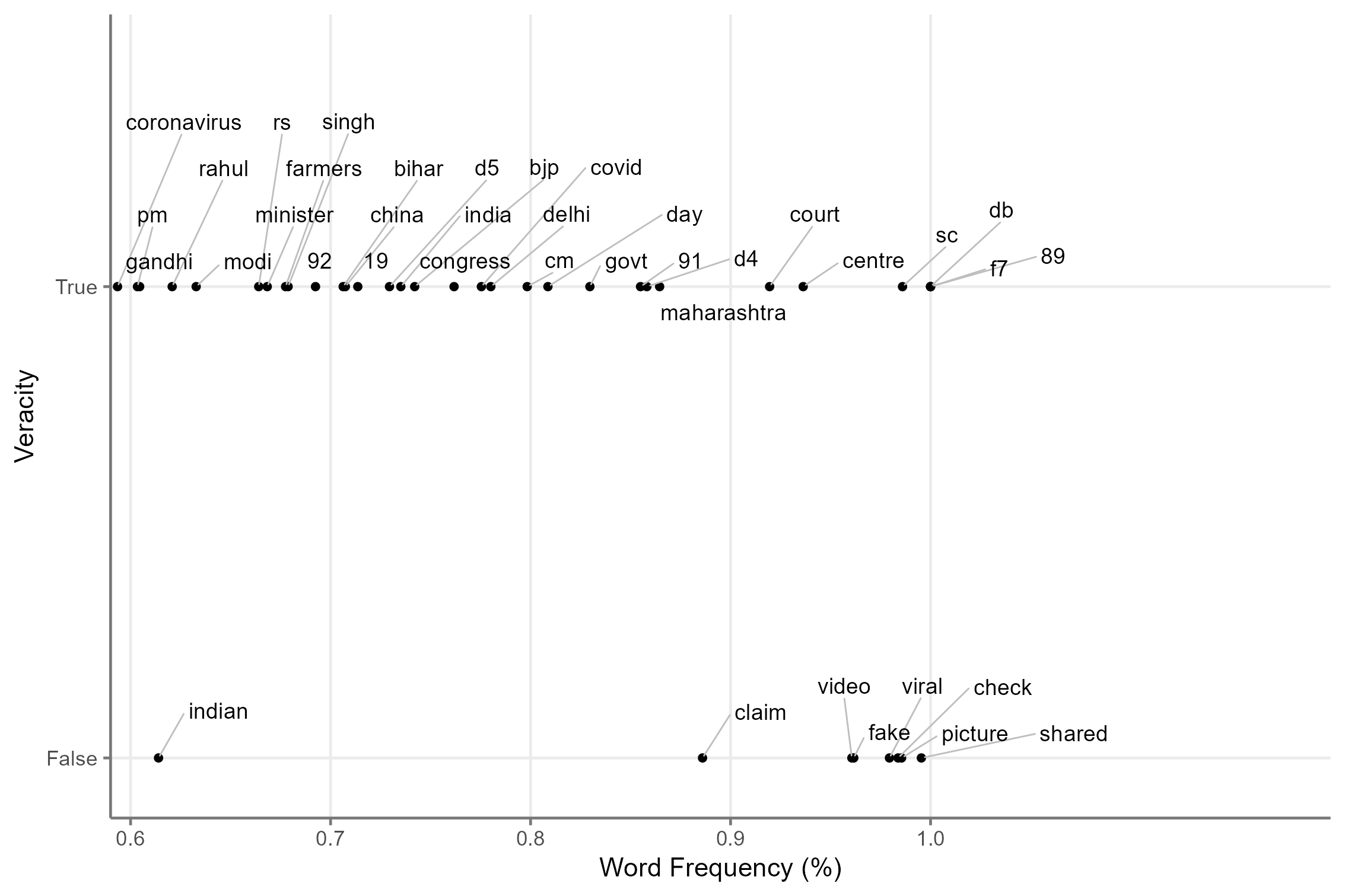}
        \caption{IFND}
        \label{fig:ifnd}
    \end{minipage}%
    \begin{minipage}{0.5\linewidth}
        \centering
        \includegraphics[width=\linewidth]{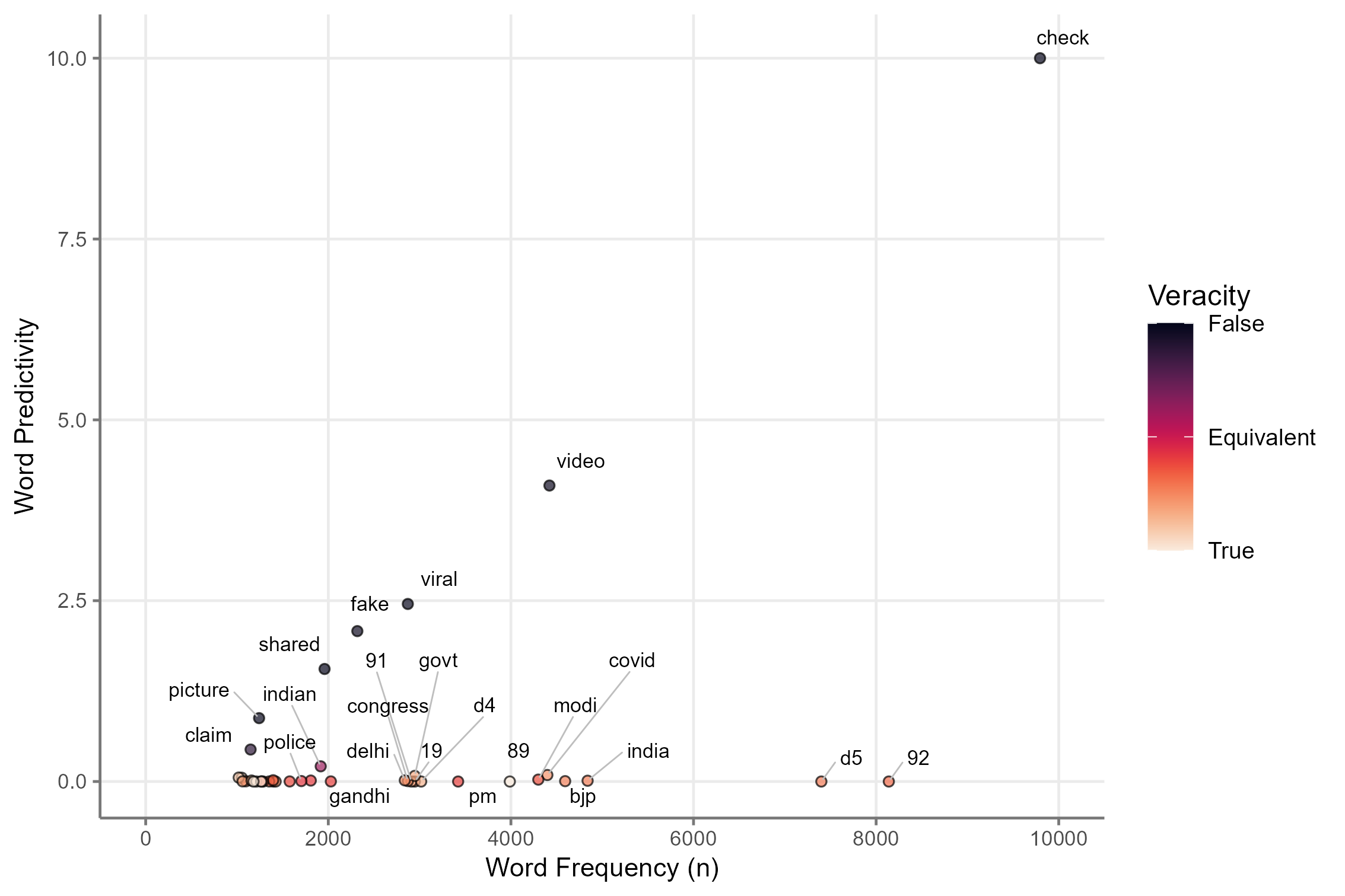}
        \caption{IFND Predictivity}
        \label{fig:pred_ifnd}
    \end{minipage}
\end{figure*}

\begin{figure*}[h!]
    \centering
    \begin{minipage}{0.5\linewidth}
        \centering
        \includegraphics[width=\linewidth]{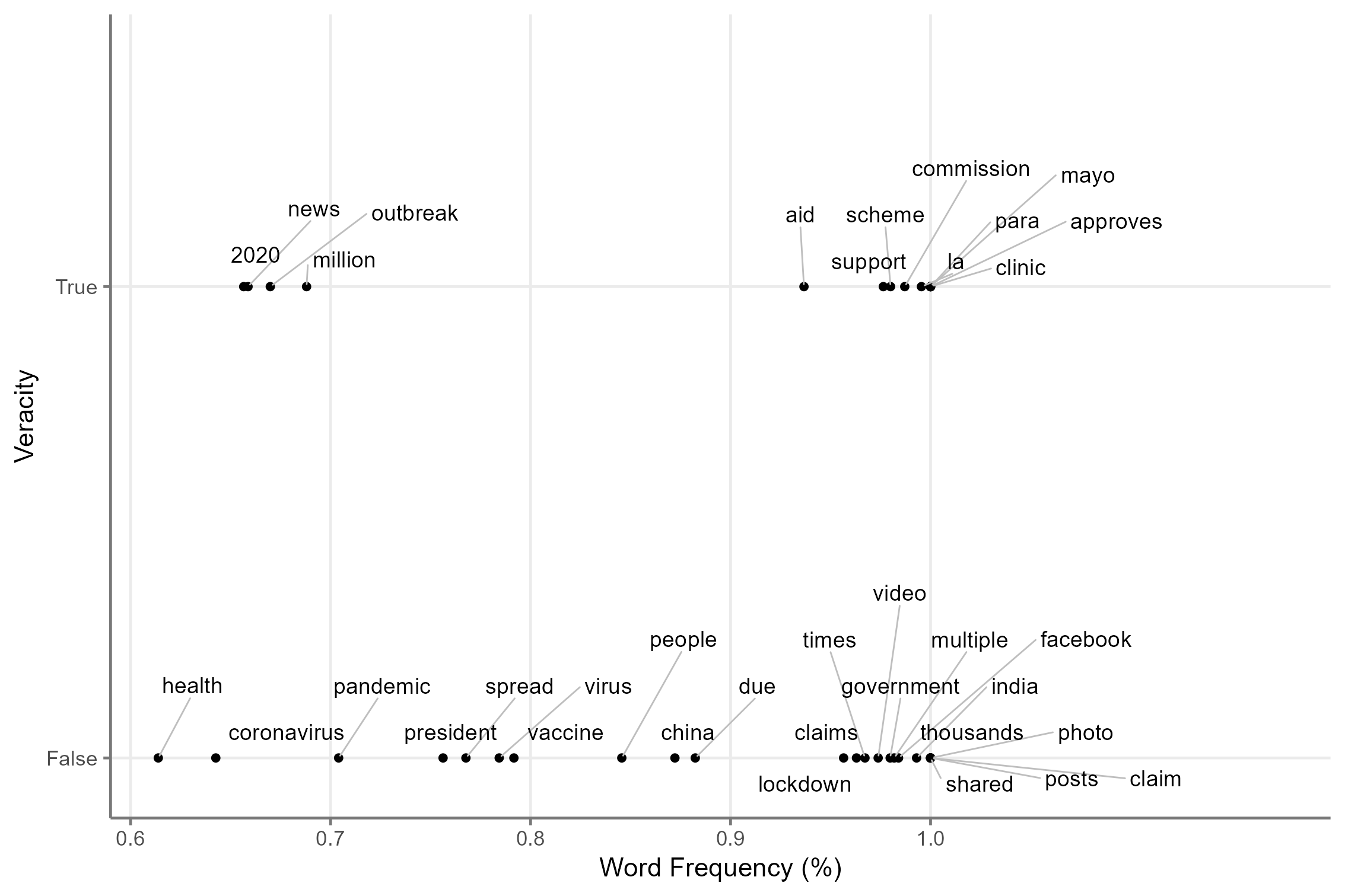}
        \caption{MM-COVID}
        \label{fig:mmcovid}
    \end{minipage}%
    \begin{minipage}{0.5\linewidth}
        \centering
        \includegraphics[width=\linewidth]{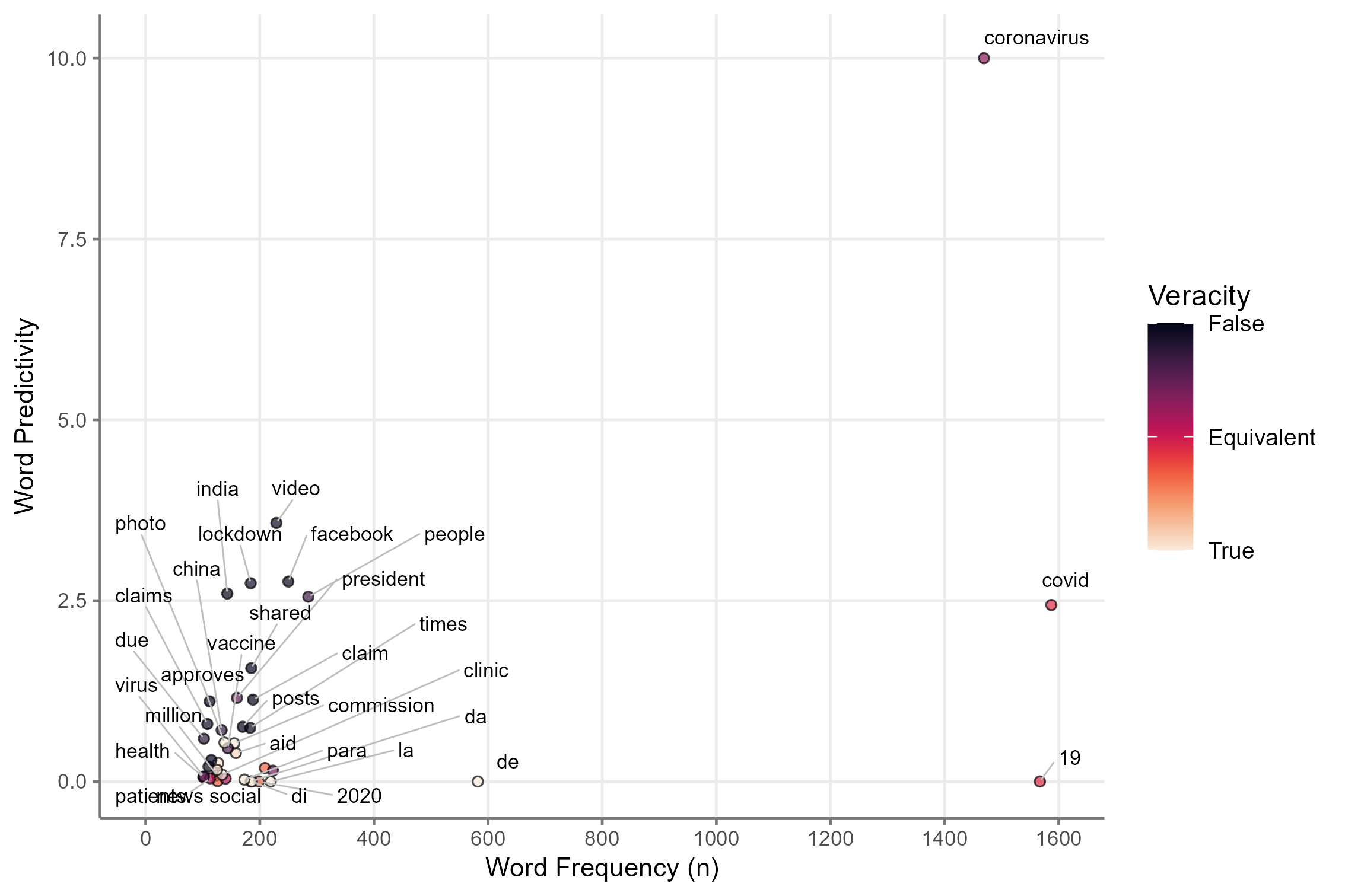}
        \caption{MM-COVID Predictivity}
        \label{fig:pred_mmcovid}
    \end{minipage}
\end{figure*}

\begin{figure*}[h!]
    \centering
    \begin{minipage}{0.5\linewidth}
        \centering
        \includegraphics[width=\linewidth]{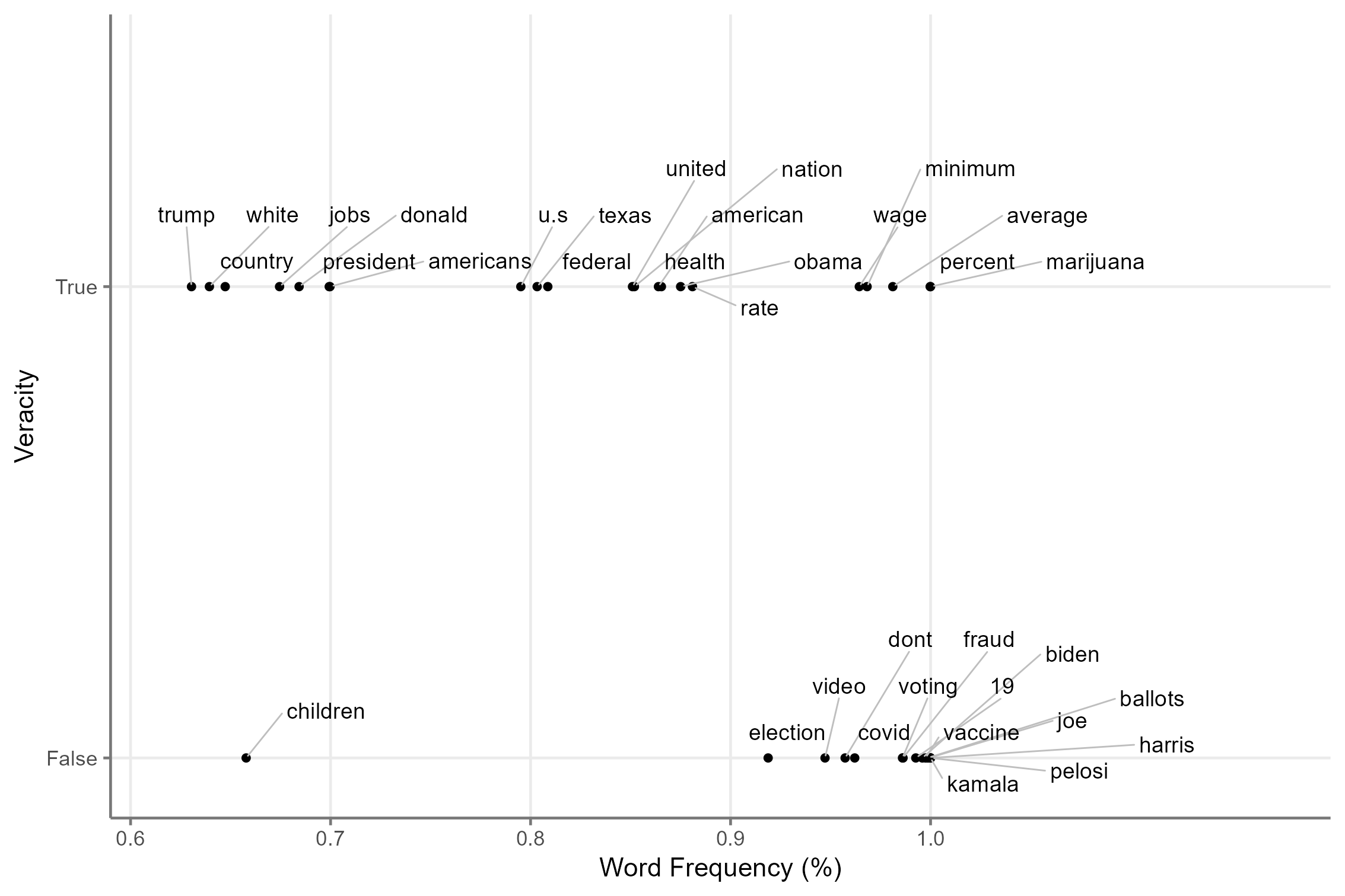}
        \caption{Truthseeker2023}
        \label{fig:truthseeker2023}
    \end{minipage}%
    \begin{minipage}{0.5\linewidth}
        \centering
        \includegraphics[width=\linewidth]{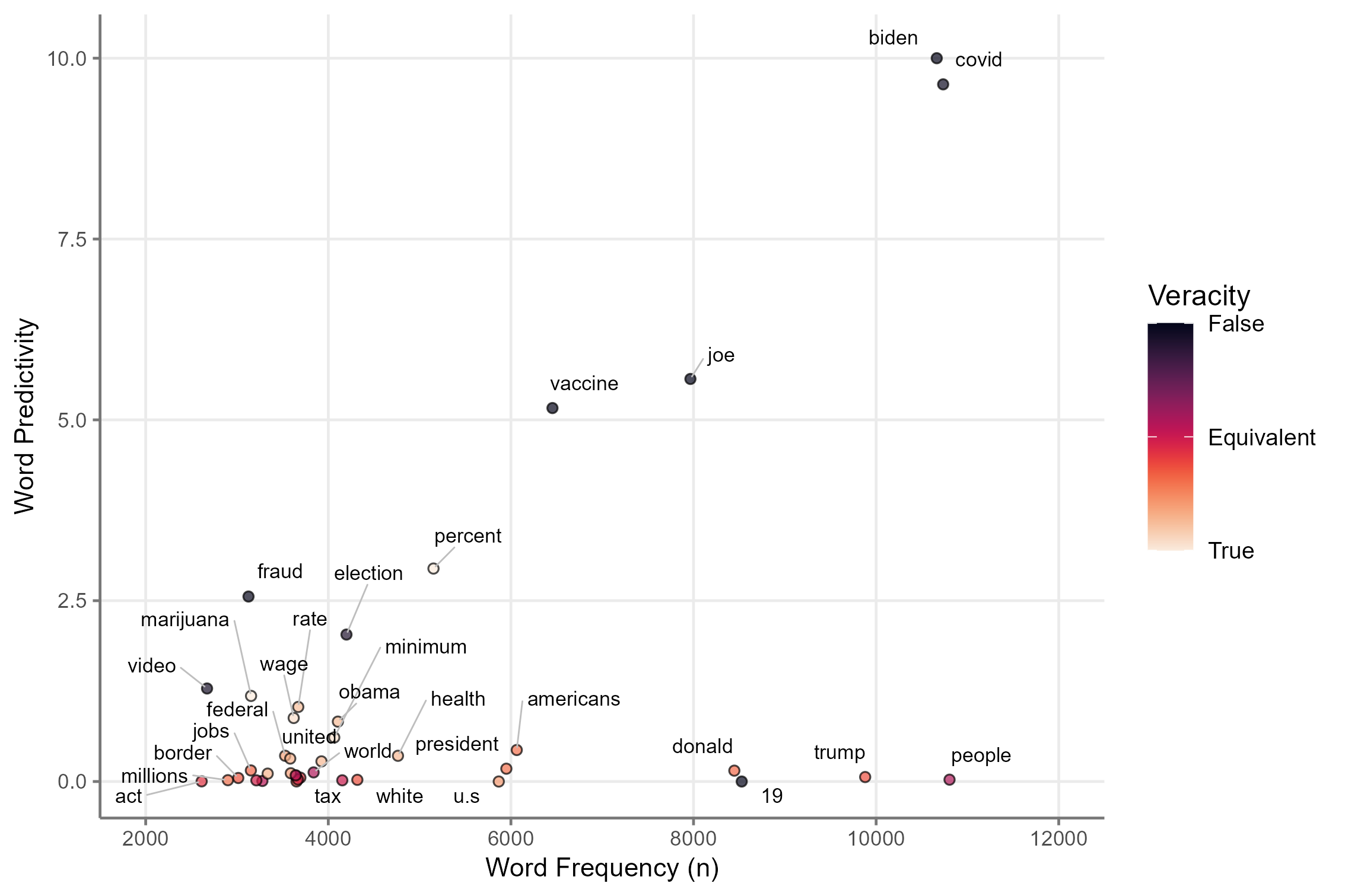}
        \caption{Truthseeker2023 Predictivity}
        \label{fig:pred_truthseeker2023}
    \end{minipage}
\end{figure*}

\begin{figure*}[h!]
    \centering
    \begin{minipage}{0.5\linewidth}
        \centering
        \includegraphics[width=\linewidth]{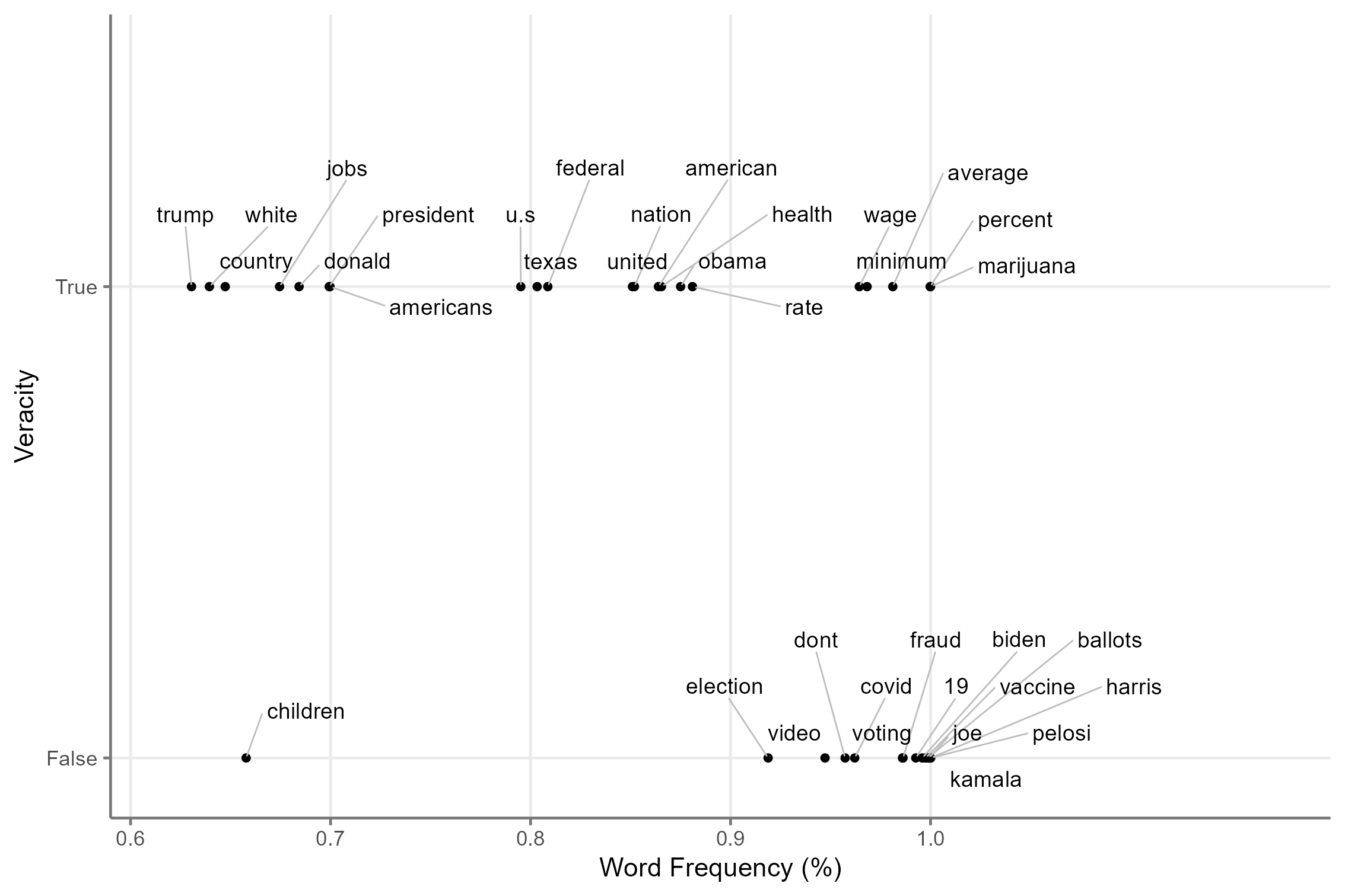}
        \caption{Twitter16}
        \label{fig:twitter16}
    \end{minipage}%
    \begin{minipage}{0.5\linewidth}
        \centering
        \includegraphics[width=\linewidth]{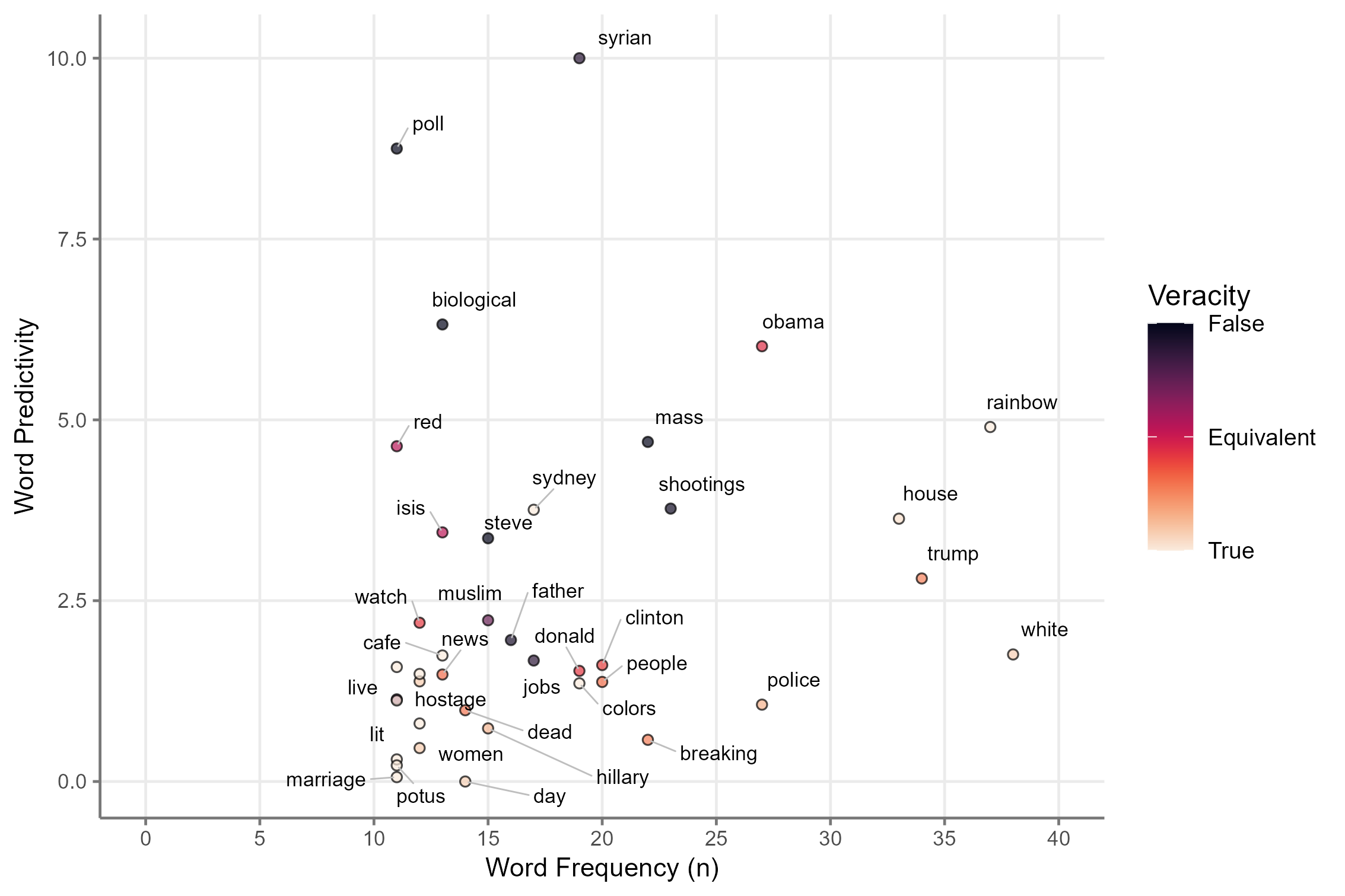}
        \caption{Twitter16 Predictivity}
        \label{fig:pred_twitter16}
    \end{minipage}
\end{figure*}

Furthermore, figures ~\ref{fig:ifnd}, ~\ref{fig:mmcovid}, ~\ref{fig:truthseeker2023} and ~\ref{fig:twitter16} show the distribution of the 40 most frequent words across the datasets IFND, MM-COVID, Truthseeker2023, and Twitter16, the four claim datasets with the highest macro F1 score (\%). The prevalence of the words in each veracity category was calculated using their relative frequency. A word positioned at x = 1 indicates that it is systematically associated with the veracity category specified by the label, meaning that 100\% of statements containing this word are labeled the same way. Additionally, figures ~\ref{fig:pred_ifnd},~\ref{fig:pred_mmcovid}, ~\ref{fig:pred_truthseeker2023}, and ~\ref{fig:pred_twitter16}, plotted using \href{https://github.com/JasonKessler/scattertext}{ScatterText}, highlight the words that had a significant impact on the random classifier presented in Table ~\ref{tab:keywords_correlations}, with color indicating the frequency of a word's association with a label.

\input{tables/keywords}

\subsection{Supplement on Temporal Analysis}
\label{app:temporal}

Table \ref{tab:temporal} presents the results of the spurious temporal correlations discussed in Section \ref{sec:data_quality}. Similar to the keywords analysis, the datasets located above the solid line include those with claims, whereas those below the line consist of paragraph datasets.

\input{tables/temporal}

\subsection{Feasibility Evaluation}
\label{app:feasibility}
\subsubsection{Claims Feasibility Evaluation}
\label{app:ai-feasibility-eval}

The prompt where it is explicitly indicated that the AI veracity assessment system has access to web search is as follows:

\begin{lstlisting}[linewidth=\columnwidth,breaklines=true]
The following statement is going to be given to an AI system to determine if it's true or false and write an explanation why.
Statement: '{statement}'

The only thing the AI will be given is the statement itself, as written above - no context, visuals, or any other information. Your task is to assess if the AI could possibly give a valid answer. Note that this is not about assessing how likely the AI is to give the right answer, but whether it's even possible to evaluate the veracity of the statement based on the information given. The AI will have access to a web search system to look for both primary and secondary sources, but the evaluation might still be impossible if there is too much ambiguity or missing context.

For example, here is a non-exhaustive list of information that might make it hard to evaluate the veracity of a statement if missing:
1. Identity of a key person, such as the speaker or someone else referenced ambiguously in the statement.
2: Location, if veracity depends on it but it isn't provided. 
3. Textual information or evidence that's mentioned in the statement but not supplied.
4. Visual or audio evidence mentioned in the statement (note that the AI will only be given the statement text). 
5. Temporal information. Note that the date the statement was made is unknown. This might not be relevant, though, if the statement could be evaluated as true or false regardless of when it was made.
6. There's no claim for which evaluating the veracity even makes sense.

Rate on the following scale how possible it seems to evaluate the veracity of the statement:

1: Feasible, assuming that the retrieval of external knowledge is possible- There is some clear ambiguity, missing context, or multiple potential interpretations. But there seems to be around one-half chance of evaluating the meaning as intended or figuring out the context from a strong knowledge base or web search.
0: Impossible to evaluate, even with access to external knowledge retrieval systems. There are clearly multiple valid ways the statement could be interpreted that would strongly influence the veracity, mandatory and irrecoverable information is missing, or the statement contains no claim or is downright nonsensical.

Give a brief explanation, then write a vertical bar "|", followed by your rating as a number alone.

\end{lstlisting}

The ``search disabled'' version, where it is not specified explicitly that the AI has access to web search, is exactly the same as above except omitting the sentence ``The AI will have access to a web search system to look for both primary and secondary sources, but the evaluation might still be impossible if there is too much ambiguity or missing context.''

We count a claim as AI-labeled ``feasible'' as long as the AI marked the claim as either feasible with search or as feasible- no search required. We use GPT-4o-mini as the AI model.

\subsubsection{Paragraphs Feasibility Evaluation}

The prompt used for feasibility evaluation on paragraphs datasets is the same as the one for shorter claims.

\subsubsection{Copy of Human Annotator Instructions}

The concept of claim feasibility is inspired by point 4 ("Methodology") of \\https://arxiv.org/abs/2401.01197. As mentioned in this paper, resolving ambiguity in a statement can be facilitated through web retrieval, particularly when context is missing but can be inferred. For example, when an unknown person is mentioned in a specific event (e.g., "politician X held a press conference about COVID on September 12, 2021 at the White House").

In other words, the feasibility label indicates whether a language model (LLM) has enough information to assess the truthfulness of a claim. To achieve this, three categories are defined:

\paragraph{Feasible}

The statement provides enough context for the LLM to determine its truthfulness with certainty.

\textbf{Example 1:} "Bill Clinton death ruled a homicide, death by poison."
\\
The statement contains all necessary information: the person (Bill Clinton) and the event (his alleged death by poisoning). Sufficient information is present to verify the claim (and we know Bill Clinton is still alive).

\textbf{Example 2:} "Ohio State scored fewer points than Purdue at the 1947 NCAA Swimming and Diving Championships."
\\
The statement includes sufficient details (institution, year, competition) to allow for factual verification without additional context.

\paragraph{Feasible with Web Search}

Some key information is missing, preventing the LLM from determining the claim’s truthfulness without retrieving additional data online.

\textbf{Example:} "A law allows people to go for a run during the state of alarm in Spain."
\\
The statement references a specific law, but it is not explicitly identified. A web search would be necessary to locate the relevant legal text and verify the claim. Since the country (Spain) and general content of the law are mentioned, this facilitates an easy online search.

\paragraph{Not Feasible}

The statement is too vague or incomplete for a web search to provide sufficient evidence for verification.

\textbf{Example:} "The (COVID-19) cases are going up but it's because the testing is going up."
\\
The statement lacks crucial details such as the time period and location, making factual verification impossible. The claim might be false if based on misinformation disseminated during the COVID-19 pandemic, or it might be true if made during a period when increased testing corresponded to rapid virus spread.

\subsubsection{Human Annotator Team and Aggregation}
\label{app:human-feasibility-annotation}

Our annotating team included 8 human experts: 3 authors and 5 colleagues of the authors. Each annotator completed an approximately equal number of examples (\~290). On the portion of the data labeled by multiple annotators (see below), the average pair-wise Cohen Kappa was 0.519 and percentage agreement 76.9\%, indicating substantial agreement.

When aggregating these annotations, we note that this setting is slightly different from typical ones where the majority vote is expected to converge to the true label. Here, a claim can be proven infeasible by counterexample: demonstrating that there are two possible contexts it could refer to. This means that a minority of annotators who think of a counterexample could be correct, even while a majority misses the ambiguity.

This creates a challenge for determining how to combine annotations from multiple annotators. In our annotation process, to obtain both examples with multiple annotators and maximize how much of each dataset we could cover, approximately 25\% was distributed to have a single annotator, 60\% to have two annotators, and the remainder had three. In the aggregation process, we first convert data that was labeled as feasible with or without search to just ``feasible''. The key question then is how to tiebreak cases where two annotators disagree.

We consider three options. First, we could tie-break in favor of ``not feasible'', setting a lower bound on overall feasibility. This may be the correct measure, by the logic above. However, we could also set a generous upper bound by tie-breaking in favor of ``feasible''. There does not seem to be clear reasoning supporting this upper bound being the true value, but it does provide a stress test for our arguments that feasibility is a significant issue. Finally, we could set a middle ground by taking the average of the two.

\begin{figure*}[h!]
    \centering
    \includegraphics[width=\textwidth]{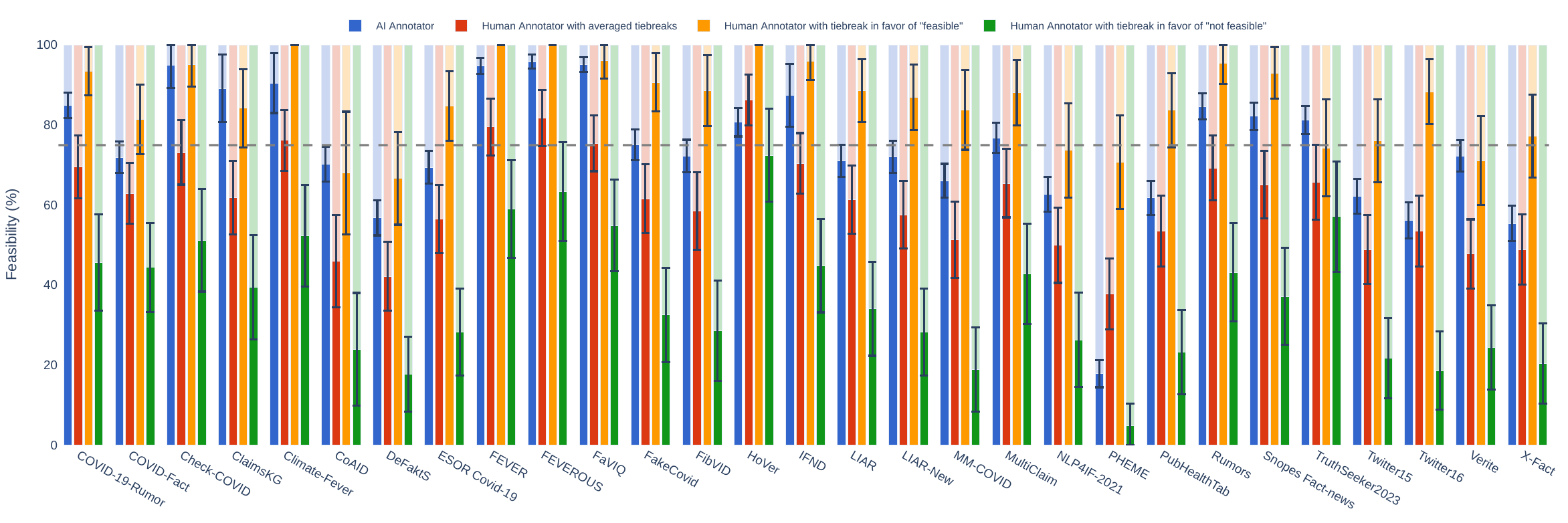}
    \caption{Comparison between human experts and automated evaluation of feasibility, aggregated based on 40 examples sampled from each data source for human annotations, and 300 for automated evaluation. }
    \label{fig:feasibility_agreements_appendix}
\end{figure*}

In Figure~\ref{fig:feasibility_agreements_appendix}, we present each of these measures. We see that even with the most generous, upper bound assessment, many datasets still have a great deal of infeasible examples. Meanwhile, the lower bound suggests many datasets could have incredibly low feasibility, to an extent that predicting veracity on the text of these datasets is not only partially but almost entirely predicting noise. In the main paper, we take the moderate approach with the average. We note, though, that even more extensive investigation of this phenomenon could be a good area for future work, and might reveal an even more severe problem than we highlight in our main paper.

\subsection{Feasibility and Predictability} 
\label{app:feasibilitypredictability}

The veracity of infeasible examples, by definition, cannot be assessed definitively -- but predictive systems can still make predictions for them, which may or may not turn out correct. In this section, we investigate empirically the connection between feasibility and correct predictions.

To do so, we sample 300 examples from the test set of every claims dataset, with the exception of six smaller datasets that have less than 300 samples in their test set, where we instead use the train set. For each of these examples, we perform veracity prediction using GPT-4o with OpenAI's web search tool, with the following prompt (this is very similar to the other search-enabled predictive results that followed \citet{tian2024web}, but a simpler and more recently available procedure for this more recent experiment):

\begin{lstlisting}[linewidth=\columnwidth,breaklines=true]
  Your task is to analyze the factuality of the given statement.

  You may invoke the "search" tool as many times as needed to retrieve up-to-date information. However, before invoking the tool, you must explain your rationale for doing so.

  After providing all your analysis steps, summarize your analysis and state "True statement; Factuality: 1" if you think the statement is factual, or "False statement; Factuality: 0" otherwise.
\end{lstlisting}

To match the GPT-4o predictive system used here, we likewise use GPT-4o to predict feasibility, following the same process described in Appendix~\ref{app:ai-feasibility-eval}). GPT-4o predicts lower feasibility rates than GPT-4o-mini used in the main paper, but the relationships described below are similar for both models.

\begin{figure*}[h!]
    \centering
    \includegraphics[width=\textwidth]{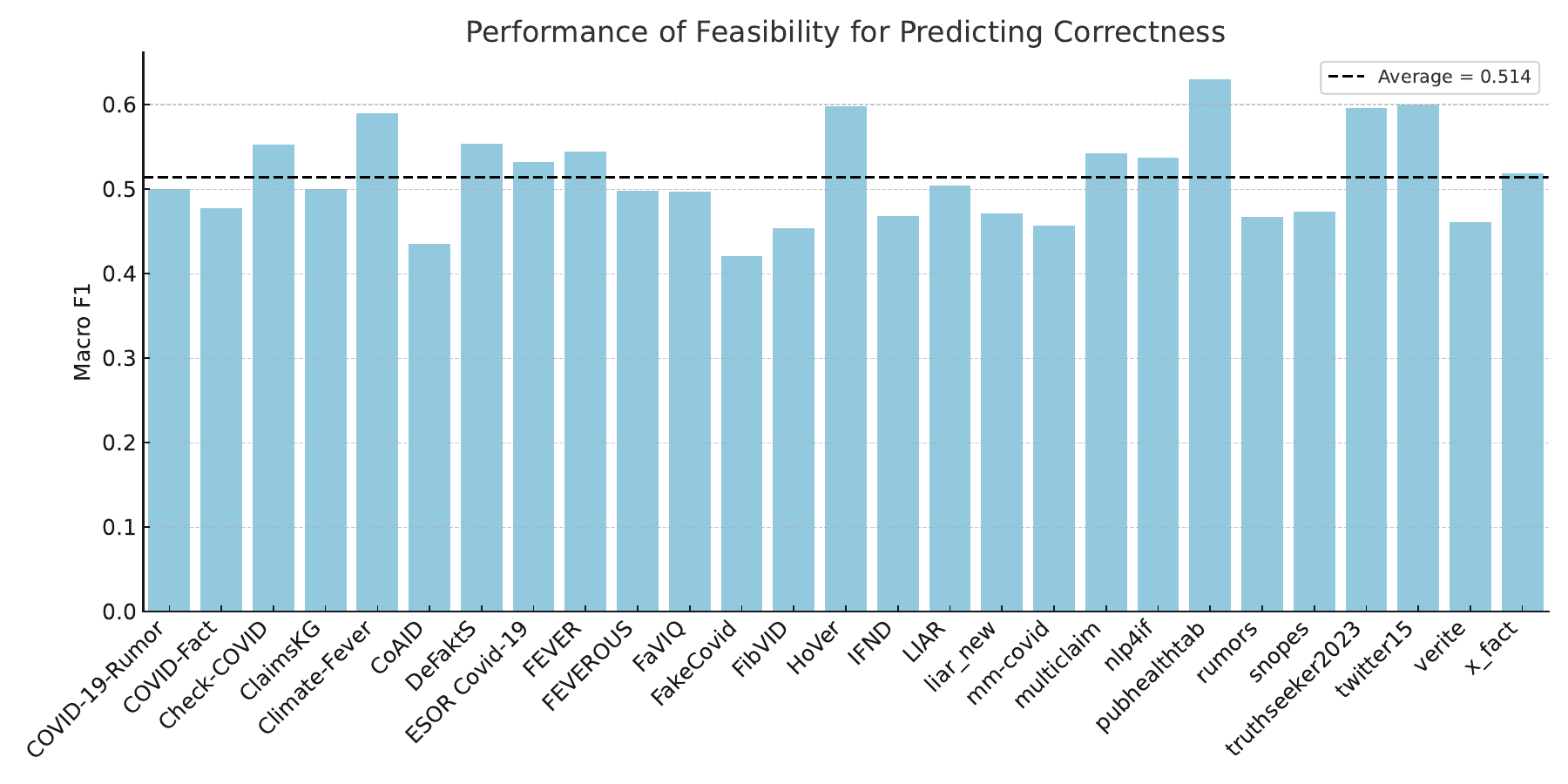}
    \caption{Performance (in macro F1 score) of feasibility at predicting whether veracity classification will be correct. Performance is limited for most datasets.}
    \label{fig:feasibility-predicting-correctness}
\end{figure*}

In Figure~\ref{fig:feasibility-predicting-correctness}, we observe a distinct but limited relationship between feasibility and predictability at the example level. Averaged across the datasets, feasibility has 51.4 macro F1 in predicting whether the classifier will evaluate veracity correctly. Meanwhile, if we pool the data, feasibility then has 56.0\% macro F1 in predicting veracity. This is above the 50\% random threshold, so there is some correspondence. At the same time, it is clearly far from a standalone explanation of classifier success or failure.

\begin{figure}[h!]
    \centering
    \includegraphics[width=\linewidth]{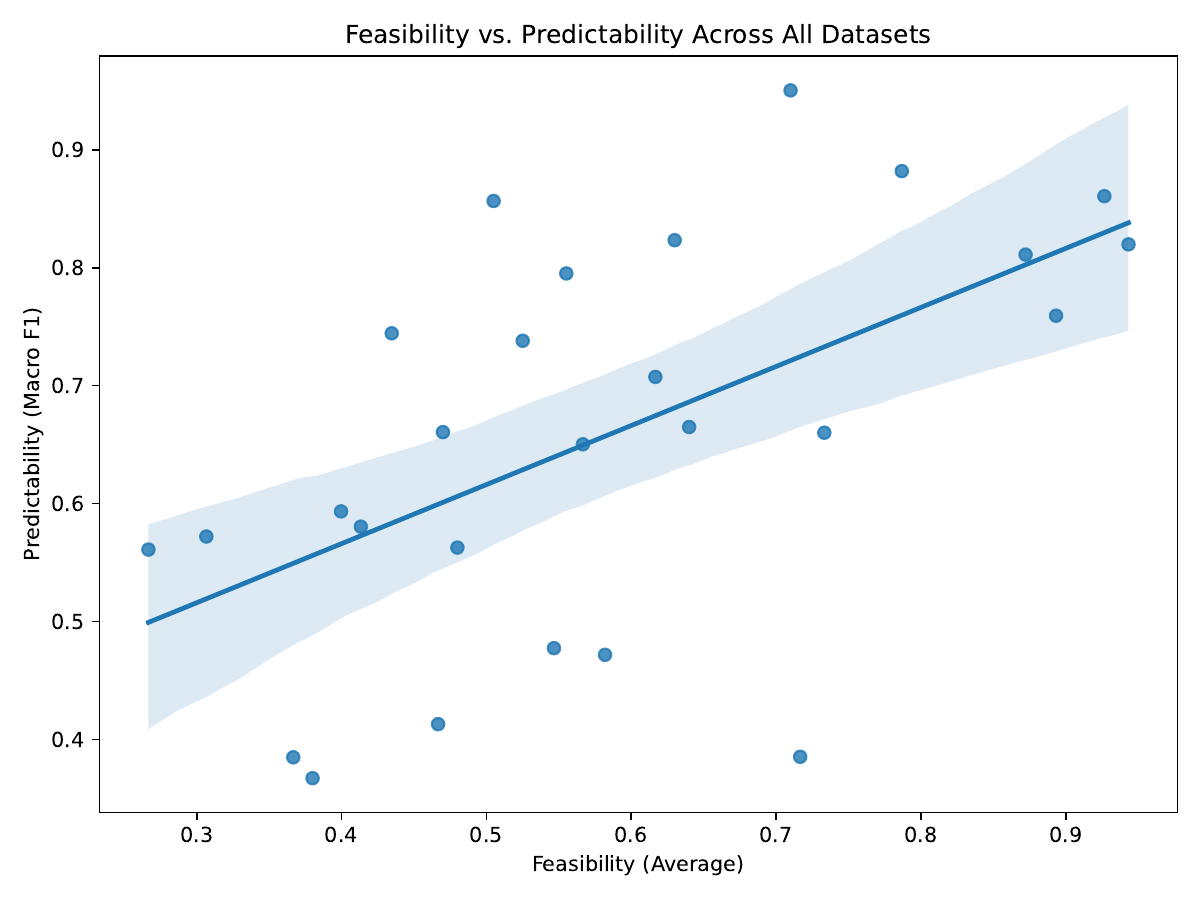}
    \caption{Feasibility vs. Predictability, with trend line plotted by OLS with 95\% confidence interval. There is a substantial positive correlation: more feasible datasets are more predictable.}
    \label{fig:feasibility_vs_predictability}
\end{figure}

However, if instead of trying to make assessments on individual examples, we compare the overall feasibility (i.e., proportion of feasible examples) and prediction correctness rates (measured by macro F1 of the predictions vs. ground truth veracity labels) of each dataset, then we observe a strong correspondence: there's 0.563 Pearson correlation, and we observe the positive relationship in Figure~\ref{fig:feasibility_vs_predictability}.

Overall, it makes sense that the example level results are noisy: one can make a correct or incorrect prediction for myriad reasons, and even if it's infeasible to \emph{reliably} make a correct one, one can get lucky in interpreting it the right way, finding the right thing in the search results, data leakage, just rolling the dice successfully on a random prediction, etc. There's also noise in the feasibility classifier, and even humans disagree around a quarter of the time on which examples are actually feasible (Appendix~\ref{app:human-feasibility-annotation}).

Aggregate feasibility, meanwhile, gives quite a strong indicator of dataset difficulty. This can mean two things: (A) There is just a lot of noise at the example level, which is averaged out at the group (dataset) level, (B) both feasibility and performance reflect some other aspects of dataset difficulty and quality: e.g., complicated claims seem less feasible and are indeed harder to assess. We hypothesize based on these results that both these possibilities are at play, in other words, there is a causal within-group effect partially obscured by noise, plus confounding at group level. We caution that many complex and intersecting factors are at play between dataset characteristics and veracity prediction systems, so this result may not be representative in all scenarios. Nonetheless, we see how feasibility can be highly informative in assessing predictive performance and dataset difficulty.

\section{Baselines and Metrics Supplement}

\subsection{Implementation Details of GPT-4 with Web Search Predictive System}
\label{app:web-search-details}

We implement our web-search predictive system by combining a state-of-the-art ``main agent'' LLM (OpenAI gpt-4-turbo-0409) with a less powerful but more efficient and cost-effective ``search agent'' LLM (Cohere command-r). We provide the search agent access to the internet through a Retrieval-Augmented Generation pipeline (RAG, implemented using the Cohere search connector\footnote{https://docs.cohere.com/docs/overview-rag-connectors}.) Specifically, the Cohere search connector applies multiple layers of filtering and reranking to efficiently condense a large number of sources from the web into a succinct response to the query from the main agent. Before any filtering was applied, the total number of tokens retrieved is usually in the range of hundred of thousands of tokens for every single example in the dataset. It would be prohibitively expensive and inefficient if all these sources need to be parsed using the gpt-4-turbo main agent. The summary that the search agent produces, which usually consists of fewer than 200 tokens, is substantially more efficient for the main agent to process while retaining most of the relevant details about the statement.

\begin{itemize}
    \item Main agent analyzes statement (chain of thought) and proposes queries, if any, to the search agent.
    \item Search agent: \begin{itemize}
        \item Find relevant documents via open web search. ($\ge 100{\rm K}$ tokens)
        \item Apply re-ranking and filtering. ($\sim 50{\rm K}$ tokens)
        \item Generate condensed response to query. ($\sim 200$ tokens) 
    \end{itemize}
    \item Main agent analyzes evidences from the search agent. Invoke search agent multiple times as needed.
    \item Main agent summarizes evidences and draw conclusion.
\end{itemize}

For further discussion of how this works, please refer to \citet{tian2024web}.

\subsection{Contradiction Between  Ground Truth Label and Predictive System}
\label{app:labelevalexamples}

The instances where labelers marked ``Predictive system is not wrong,'' even though the system's output contradicted the ground truth label, can be attributed to differences in timing, interpretation, or problems with the ground truth labels and the claims themselves.

Different timing may lead to contradictions. For instance, in the MM-Covid dataset, there was a claim stating, ``Lysol disinfectant label says it was tested against the new coronavirus.'' The AFP Fact Check labeled this claim as false in September 2020 because, at the time, no Lysol product had been tested against COVID-19. However, a Lysol product was later developed and tested, leading the predictive system to label the claim as true. Similarly, in the LIAR dataset, a claim that ``Inflation has gone up every month of the Biden presidency and just hit another 40-year high'' was rated as mostly true by PolitiFact in April 2022. However, when the predictive system analyzed the claim using data from January 2024, it labeled it as false, correctly accounting for more recent information.

Another source of contradiction can be the interpretation of the claims. One instance is this claim from the FEVER dataset: ``Dakota Fanning is not a model.'' The ground truth label was false, considering that Dakota Fanning is primarily an actress. However, the predictive system labeled it as true, considering she has engaged in modeling and has appeared in various magazine photoshoots. Here, the system's broader interpretation of what constitutes a ``model'' led to a contradiction, yet it is not necessarily wrong.

Contradictions also arise due to the specific wording of claims, which is especially prevalent in the MM-Covid dataset. For instance, the ground truth label marked the claim ``President Donald Trump's statement that lupus patients are not vulnerable to COVID-19 is not true'' as false, focusing solely on Trump's statement. However, the predictive system, which analyzed the entire sentence, classified it as true. The predictive system explained that lupus patients are vulnerable to COVID-19, and thus Donald Trump's statement is indeed not true. Another example is the claim, ``These are 6 of the main differences between flu and coronavirus,'' which had a ground truth label of true based on a headline from the MIT Technology Review. The predictive system, however, labeled it as false, arguing that the differences between the flu and coronavirus cannot be strictly limited to six. The problem is not the labelling  of the predictive system, rather the ground truth labels and the claims themselves. 

\subsection{Supplement on Manual Labeling of Prediction Validity}
\label{app:manuallabel}

\paragraph{LIAR-New} Two authors labeled 100 samples that the GPT-4 (with web search) predictive system got wrong according to standard comparison with the ground truth labels from the professional fact-checkers at PolitiFact (which the dataset is sourced from). The labelers considered the input statement, the reasoning of the predictive system, and the PolitiFact fact-checking article. They each labeled every example, with a 3-way schema: ``Predictive system is wrong'', ``Uncertain / open to interpretation'', ``Predictive system is not wrong''.

This led to 0.36 Cohen Kappa agreement and 60\% percentage agreement. The agreement cases within these results indicated a large number of cases where the predictive system was not wrong---38 out of 60 examples where the labels agreed---but to further reinforce the validity of the labeling, the annotators discussed each disagreement and produced a single resolution label. In this final result, of the 100 cases, 30 were ``Predictive system is wrong'', 15 were ``Uncertain / open to interpretation'', and the remaining 55 were ``Predictive system is not wrong''.

The two annotators also manually labeled 100 examples that were originally marked correct. They agreed 76 were not wrong, 2 were uncertain, and 1 was wrong. There were only 5 additional examples that were marked wrong by one but not both annotators. 

\paragraph{FEVER} The same two authors then labeled predictions based on GPT-3.5 (with web search) on the FEVER dataset that were  marked incorrect by standard categorical label comparison. Here, there is no fact-checking article to reference, so the authors looked up any necessary information themselves, again seeking to determine if the LLM's explanation was correct. First, they labeled 10 examples together to synchronize the labeling process, then both labeled the same 100 independently. We discard the first 10. On the 100, the labels had 0.51 Cohen Kappa agreement score and 70\% agreement. Since the initial agreement was higher, we did not conduct a resolution process on this data. 38 examples were marked ``Predictive system is not wrong'' by both labelers, and 56 by at least one.

\paragraph{MM-COVID}

Again, the annotators labeled examples that the categorical labels marked incorrect. These were from the GPT-4 (with web search) version of the baseline system. There were only 70 of these total, so the annotators labeled all 70. They had 44\% agreement, and agreed that 39 examples were not wrong while agreeing only a mere 3 were wrong. An additional 25 were marked not wrong by one annotator.

Then the annotators labeled 100 examples that were correct according to the categorical labels. They agreed that 89 examples were not wrong, and there were 0 that they agreed were wrong. There were another 5 examples that were marked wrong by one but not both annotators.

\subsection{Supplement on Contradiction Evaluator}
\label{app:contradictioneval}

We implement experiments on the explanations from the GPT-4 with web search predictive system. For comparison with human labels, we use the final version after resolution described above, and drop all ``Uncertain / open to interpretation'' cases. For all versions of the evaluator, we use GPT-4-Turbo-0409, with temperature 0.0 to reduce variation. There is nonetheless some variation; to further stabilize the estimates, we ran 5 runs and report results using the mean (in the score case) or majority vote (in the binary and trinary cases).

In addition to the score-based version described in the main text, we tested binary and trinary versions of the evaluator. The score-based prompt is:

\begin{lstlisting}[linewidth=\columnwidth,breaklines=true]
In the following, you will be provided a statement and two assessments of its veracity. Your task is to evaluate if the assessments contradict each other. Note that not having all of the same evidence or content, or even reaching a different conclusion, does not alone constitute a contradiction, especially though not exclusively if they are interpreting the statement differently, or considering different time periods or other contexts. There's only a contradiction if they actually say opposing things that are not up to reasonable interpretation or context differences. 

Statement: <statement>

Assessment 1: <article>

Assessment 2: <prediction>

Now that you've ready the statement and assessments, rate how much the assessments contradict or not on a scale from 0 (no contradiction) to 10 (complete contradiction). However, you must not state your score until you've presented a concise analysis. Do not begin your response with a number. First write your analysis, then write a vertical bar "|", then finally state your contradiction score.
\end{lstlisting}

Leaving the rest of the prompt unchanged, we adjust the last paragraph as follows to get the binary version:

\begin{lstlisting}[linewidth=\columnwidth,breaklines=true]
Now that you've ready the statement and assessments, answer if the assessments contradict or not. However, you must not state your decision until you've presented a concise analysis. Do not begin your response with a label. First write your analysis, then write a vertical bar "|", then finally "1: contradiction" or "0: no contradiction".
\end{lstlisting}

And trinary:

\begin{lstlisting}[linewidth=\columnwidth,breaklines=true]
Now that you've ready the statement and assessments, answer if the assessments contradict or not. However, you must not state your decision until you've presented a concise analysis. Do not begin your response with a label. First write your analysis, then write a vertical bar "|", then finally "1: contradiction" or "0: no contradiction", or if you are not sure write "-1: unsure".
\end{lstlisting}

Binary agrees 68\% of the time with the human labels, trinary 67\% of the time, and the original score-based approach 68\% of the time. Thus, there is little difference in efficacy. We note that the trinary approach, although explicitly given the option to output ``unsure'', never used it.

%% file: tables/71_data_description.tex
\begin{table*}[h!]
\caption{Characterizing 75 common misinformation datasets. Datasets are ordered by modality, then date, and topic.}
\label{tab:datasets-summary}
\centering
\fontsize{5}{6}\selectfont
\resizebox{.99\linewidth}{!}{
\begin{tabular}[t]{lrllllrr}
\toprule

Dataset & Size & Modality & Topic & Geographic region & Language & Time start & Time end \\

\midrule
\href{https://doi.org/10.1016/j.puhe.2021.11.022}{AntiVax} & 15,465,687 & Claims & Health & USA & EN & 01/12/2021 & 31/07/2021 \\
\href{https://arxiv.org/abs/2006.00885}{CoAID} & 301,177 & Claims & Covid-19 & - & EN & 01/12/2019 & 01/09/2020 \\
\href{https://ieeexplore.ieee.org/document/9377956/}{Counter-covid-19-misinformation} & 155,468 & Claims & Covid-19 & International & EN & 21/01/2020 & 20/05/2020 \\
\href{https://doi.org/10.3389/fpsyg.2021.644801}{COVID-19-Rumor} & 7,179 & Claims & Covid-19 & - & EN & 01/2020 & 03/2020 \\
\href{https://arxiv.org/abs/2005.00033}{Covid-19-disinformation} & 16,000 & Claims & Covid-19 & International & 4 lang. & 01/2020 & 03/2021 \\
\href{https://arxiv.org/pdf/2311.18195.pdf}{Covid-vaccine-misinfo-MIC} & 5,952 & Claims & Covid-19 & Brazil, Indonesia, Nigeria & EN, PT, ID & 2020 & 2022 \\
\href{https://esoc.princeton.edu/publications/localized-misinformation-global-pandemic-report-covid-19-narratives-around-world}{ESOC Covid-19} & 5,613 & Claims & Covid-19 & International & 35 lang. & 01/01/2020 & 1/12/2020 \\
\href{https://arxiv.org/abs/2006.11343}{FakeCovid} & 7,623 & Claims & Covid-19 & International & 40 lang. & 04/01/2020 & 01/07/2020 \\
\href{https://doi.org/10.1016/j.tele.2021.101688}{FibVID} & 1,353 & Claims & Covid-19 & International & EN & 02/2020 & 01/2021 \\
\href{https://doi.org/10.1145/3472720.3483617}{WICO} & 364,325 & Claims & Covid-19 & - & EN & 17/01/2020 & 30/06/2021 \\
\href{https://aclanthology.org/P17-1066/}{Twitter16} & 818 & Claims & Various & - & EN & 03/2015 & 12/2016 \\
\href{https://doi.org/10.14778/3329772.3329778}{Rumors} & 1,022 & Claims & Various & USA, UK, China, India & EN & 01/05/2017 & 01/11/2017 \\
\href{https://link.springer.com/chapter/10.1007/978-3-030-30796-7_20}{ClaimsKG} & 74,066 & Claims \& Knowledge Graph & Various & International & EN & 1996 & 2023 \\
\href{https://doi.org/10.1007/s40747-021-00552-1}{IFND} & 56,868 & Claims \& Images & Various & India & EN & 2013 & 2021 \\
\href{https://link.springer.com/article/10.1007/s13735-023-00312-6}{Verite} & 1,001 & Claims \& Images & Various & International & EN & 01/01/2001 & 01/01/2023 \\
\href{https://aclanthology.org/P17-2067/}{LIAR} & 12,836 & Claims & Politics & USA & EN & 2007 & 2016 \\
\href{https://arxiv.org/pdf/2305.14928}{LIAR-New} & 1,957 & Claims & Politics & USA & EN \& FR & 10/2021 & 11/2022 \\
\href{https://www.techrxiv.org/articles/preprint/TruthSeeker_The_Largest_Social_Media_Ground-Truth_Dataset_for_Real_Fake_Content/22795130}{TruthSeeker2023} & 180,000 & Claims & Politics & USA & EN & 2009 & 2022 \\
\href{https://aclanthology.org/2023.findings-acl.888.pdf}{Check-COVID} & 1,504 & Claims & Covid-19 & - & EN & - & - \\
\href{https://storage.googleapis.com/gweb-research2023-media/pubtools/pdf/3cb6fa24437bd77f337443c05b2fdd1beded875b.pdf}{Climate-Fever} & 7,675 & Claims & Climate & - & EN & - & -\\
\href{https://ceur-ws.org/Vol-2699/paper40.pdf}{CMU-MisCOV19} & 4,573 & Claims & Covid-19 & - & EN & - & - \\
\href{https://aclanthology.org/2021.acl-long.165.pdf}{COVID-Fact} & 4,086 & Claims & Covid-19 & - & EN & - & - \\
\href{https://aclanthology.org/2022.acl-long.354.pdf}{FaVIQ} & 188,376 & Claims & Various & - & EN & - & - \\
\href{https://aclanthology.org/N18-1074}{FEVER} & 185,445 & Claims & Various & USA & EN & - &  - \\
\href{https://aclanthology.org/2021.fever-1.1.pdf}{FEVEROUS} & 87,026 & Claims & Various & - & EN & - & - \\
\href{https://aclanthology.org/2020.findings-emnlp.309.pdf}{HoVer} & 26,171 & Claims & Various & - & EN & - & - \\
\href{https://doi.org/10.1007/s13735-017-0143-x}{MediaEval}  & 15,629 & Claims \& Images & Various & - & EN \& ES & - & - \\
\href{https://arxiv.org/abs/2011.04088}{MM-COVID} & 11,173 & Claims & Covid-19 & International & 6 lang. & - & - \\
\href{https://arxiv.org/abs/1806.03713}{PHEME} & 62,445 & Claims & Newsworthy events & International & EN & - & - \\
\href{https://aclanthology.org/2022.findings-naacl.1/}{PubHealthTab} & 1,942 & Claims & Health & North America & EN & - & - \\
\href{https://www.kaggle.com/datasets/ambityga/snopes-factnews-data}{Snopes Fact-news} & 4,550 & Claims & Various & USA & EN & - & - \\
\href{https://aclanthology.org/P17-1066/}{Twitter15} & 1,490 & Claims & Various & - & EN & - & -\\
\href{https://aclanthology.org/2021.acl-short.86.pdf}{X-Fact} & 31,189 & Claims & Various & - & 25 lang.  & - & - \\
\href{https://aclanthology.org/2024.lrec-main.409.pdf}{DeFaktS} & 105,855 & Claims & Various & Germany & DE & - & - \\
\href{https://aclanthology.org/2023.emnlp-main.1027.pdf}{MultiClaim} & 31,305 & Claims & Various & International & 39 lang. & - & - \\
\href{https://arxiv.org/pdf/2109.12986}{NLP4IF-2021} & 3,172 & Claims & Covid-19 & International & AR, BUL \& EN & - & - \\
\href{https://arxiv.org/abs/1703.09398}{Benjamin Political News}  & 296 & News articles & Election & USA & EN & 02/2016 & 11/2016 \\
\href{https://www.buzzfeednews.com/article/craigsilverman/partisan-fb-pages-analysis}{BuzzFeedNews} & 2,282 & News articles & Election & USA & EN & 19/09/2016 & 27/09/2016 \\
\href{https://ceur-ws.org/Vol-3180/paper-30.pdf}{CT-FAN} & 2462 & News articles & Various & Germany, USA, Canada & EN \& DE & 2010 & 2022 \\
\href{https://arxiv.org/pdf/2312.03750.pdf}{Fake News Elections} & 38,333 & News articles & Politics & USA & EN & 04/2023 & 10/2023 \\
\href{https://doi.org/10.1145/3201064.3201100}{FakeNews} & 486 & News articles & Politics & USA & EN & 01/2016 & 10/2017 \\
\href{https://doi.org/10.1609/icwsm.v13i01.3254}{FA-KES} & 804 & News articles & Syrian war & Syria & EN & 2011 & 2018 \\
\href{https://aclanthology.org/2021.fever-1.9/}{FANG-COVID} & 41,242 & News articles & Covid-19 & Germany & DE & 02/2020 & 03/2021 \\
\href{https://doi.org/10.1002/spy2.9}{ISOT Fake News} & 44,898 & News articles & Politics & International & EN & 2016 & 2017 \\
\href{https://doi.org/10.1371/journal.pone.0227821}{Italian disinformation} & 16,867 & News articles \& Tweets & Election & Italy & EN \& IT & 01/01/2019 & 27/05/2019 \\
\href{https://arxiv.org/pdf/2306.08871.pdf}{Med-MMHL} & 40,601 & News, tweets, images \& LLM-generated & Health & USA & EN & 01/01/2022 & 01/05/2023 \\
\href{https://arxiv.org/abs/2102.04567}{NELA-GT-2020} & 1,779,127 & News articles \& Tweets & Various & USA & EN & 01/01/2020 & 31/12/2020 \\
\href{https://doi.org/10.1145/3340531.3412880}{ReCOVery} & 142,849 & News articles \& Tweets & Covid-19 & - & EN & 01/2020 & 05/2020 \\
\href{https://arxiv.org/pdf/1806.00749.pdf}{Spanish Fake News Corpus} & 572 & News articles \& Social media posts & Various & International & ES & 01/11/2020 & 31/03/2021\\
\href{https://doi.org/10.1145/3459637.3482139}{Weibo21} & 9,128 & News articles & Various & China & CN & 12/2014 & 03/2021 \\
\href{https://aclanthology.org/2020.lrec-1.349/}{BanFakeNews} & 50,000 & News articles & Various & Bangladesh & EN & - & - \\
\href{https://aclanthology.org/C18-1287}{Celebrity} & 500 & News articles & Celebrities & - & EN & - & - \\
\href{https://arxiv.org/abs/2011.03327}{COVID-19 Fake News} & 10,700 & News articles \& Social media posts & Covid-19 & - & EN & - & - \\
\href{https://arxiv.org/pdf/2202.07094.pdf}{Fact-check-tweet} & 13,070 & News articles \& Tweets & Various & International & 4 languages & - & - \\
\href{https://doi.org/10.1609/icwsm.v14i1.7350}{FakeHealth} & 440,870 & News articles \& Social media posts & Health & USA & EN & - & - \\
\href{https://aclanthology.org/C18-1287}{FakeNewsAMT} & 480 & News articles & Various & - & EN & -& - \\
\href{https://github.com/several27/FakeNewsCorpus}{FakeNewsCorpus} & 9,408,908 & News articles & Various & - & EN & - & -\\
\href{https://arxiv.org/abs/1809.01286}{FakeNewsNet} & 23,196 & News articles & Politics \& Celebrities & - & EN & - & - \\
\href{https://arxiv.org/abs/1707.03264}{FNC-1} & 49,972 & News articles & Various & - & EN & - & -\\
\href{10.18653/v1/2022.acl-long.222}{Misinfo Reaction Frames} & 25,100 & News articles & Global crises & International & EN & - & - \\
\href{https://doi.org/10.1145/3477495.3531744}{MuMin} & 21,565,018 & News articles, Tweets \& Images & Various & International & 41 languages & - & - \\
\href{http://dx.doi.org/10.3233/JIFS-179034}{TI-CNN} & 20,015 & News articles & Politics & USA & EN & - & - \\
\href{https://doi.org/10.1609/icwsm.v12i1.14985}{BuzzFace} & 1,176,713 & Social Media posts & Election & USA & EN & 01/09/2016 & 30/09/2016\\
\href{https://arxiv.org/abs/1704.07506}{FacebookHoax} & 15,500 & Social Media posts & Hoaxes & - & EN & 01/07/2016 & 31/12/2016 \\
\href{https://arxiv.org/abs/2205.06181}{FACTOID} & 3,354,450 & Social Media posts & Politics & USA & EN & 01/2020 & 04/2021 \\
\href{https://arxiv.org/abs/1911.03854}{Fakeddit} & 1,063,106 & Social Media posts \& Images & Various & - & EN & 19/03/2008 & 24/10/2019\\
\href{https://ojs.aaai.org/index.php/ICWSM/article/view/18113/17916}{VoterFraud2020} & 7,600,000 &  Social Media posts, Images \& Videos & Election & USA & EN & 23/10/2020 & 16/12/2020 \\
\href{https://dl.acm.org/doi/pdf/10.1145/3539618.3591896}{MR2} & 14,700 &  Social Media posts \& Images & Rumor & USA \& China & EN \& CN & - &  - \\
\href{https://doi.org/10.1145/3340531.3417463}{Reddit} & 12,597 &  Social Media posts & Various & USA & EN & - & - \\
\href{https://doi.org/10.1007/978-3-540-76298-0_52}{DBpedia} & 1,950,000 & Wikipedia text & Various & International & 14 lang. & - & - \\
\href{https://doi.org/10.37016/mr-2020-030}{ICWSM} & 2,500 & Images & Election & Brazil \& India & 10 lang. & 10/2018 & 06/2019 \\
\href{https://arxiv.org/abs/1901.08971}{FaceForensics++} & 1,800,000 & Images & Deepfakes & - & - & - & - \\
\href{https://doi.org/10.1108/OIR-03-2018-0101}{FCV-2018} & 380 & Videos & Various & International & 5 lang. & 2016 & 2017 \\
\href{https://arxiv.org/abs/1909.12962}{Celeb-DF} & 5,639 & Videos & Celebrities & - & - & - & - \\
\href{https://arxiv.org/abs/1812.08685}{DEEPFAKETIMIT} & 640 & Videos & Various & - & - & - & - \\
\bottomrule
\end{tabular} }
\end{table*}

%% file: 030survey.tex
\section{Survey and Construction of \cdlmd}
\label{sec:Survey}

In this section, we provide an expanded discussion of our collection of combined datasets on true and false information, including a total of 75 datasets, with 36 specifically focused on claims and statements, and 9 specifically focused on paragraphs. The full dataset contains a total of 120,901,495 observations, while the subset that we further analyze includes 1,741,146 observations. These data encompass a wide range of topics, including political issues, health concerns, and environmental questions, often related to the United States but also covering international news, headlines and online posts. The original labels within the datasets were assigned through a combination of expert evaluations and algorithmic methods. The following section provides a detailed summary of the data collection process and the characteristics of these datasets.

\input{020relatedwork}

\subsection{Collection Process}
Our data collection process involved an exhaustive search of journal and conference articles to identify relevant datasets. To achieve this, we used the Google Scholar search engine with keywords such as ``fake news'', ``disinformation'', ``misinformation'', ``dataset'', ``detection'', ``survey'', and others highlighted in Appendix~\ref{app:datacollection}. We focused on articles published between 2016 and 2024. This initial phase allowed us to collect 28 datasets.

We then expanded our selection by rigorously examining the citations in scientific articles related to these initial sources. Some articles listed available datasets in their literature reviews or surveys, enabling us to incorporate additional data. Through these combined approaches---which are further detailed in the Appendix---we identified 75 publicly accessible datasets, presented in Table~\ref{tab:datasets-summary}.

For our analyses, we refined our selection to focus on claims datasets and paragraph datasets. Claims datasets contain textual claims, defined here as short statements ranging from one to two sentences. Tweets are included in this definition, while lengthier online and social media posts are excluded. Paragraph datasets contain more comprehensive arguments, defined by having more than two sentences. Articles and social media posts fit under this definition. Our selection excludes hybrid datasets, which are defined as possessing data types which can be classified as both claims and paragraphs. These two dataset types are qualitatively different because statements and claims are more concise than paragraphs, such as OP-ED and news articles, which often include opinions, commentary, and contextual details. This extraneous information can potentially obscure the core claim or statement and introduce noise in the labeling process, as information can be partly true or false.

\subsection{Supplement on Data Collection}
\label{app:datacollection}

For the dataset search, all of the following set of words, as well as the ones presented in Section 3.1, were used in Google Scholar: ``fake news'', ``false news'', ``fake news dataset'', ``false news dataset'', ``fake news database'', ``false news database'', ``misinformation dataset'', ``misinformation database'', ``misinformation detection'', ``misinformation survey'', ``disinformation dataset'', ``disinformation database'', ``disinformation detection'', ``disinformation survey'', ``fact check dataset'', ``fact check database'', ``benchmark for fake news detection'', ``benchmark dataset for fake news'', ``misinformation data'', ``dataset for evidence-based fact-checking'', ``fact-checking corpus'', ``fact verification corpus'', and ``misinformation detection review''. In addition to these terms, and as mentioned previously, a year filter was used using the advanced search. Only articles dated between 2016 and 2024 were included.

Once the initial datasets were identified, we then expanded our selection by (1) identifying the most frequently cited papers related to these datasets (based on the number of citations in Google Scholar) and (2) carefully reviewing these papers to uncover additional dataset. This review process primarily focused on analyzing the articles' literature reviews and reference lists to identify datasets that were mentioned and could be pertinent to our research. For instance, according to Google Scholar, the article by \citet{shu2020fakenewsnet}, which introduces the FakeNewsNet dataset, has been cited 1,190 times, ranking it among the top four most frequently cited articles that we collected. Based on this, we proceeded to review \textit{Section 2} of the article, titled \textit{Background and Related Work}. In this section, the authors mention six existing datasets for misinformation detection: BuzzFeedNews, LIAR, BS Detector, Credbank, BuzzFace, and FacebookHoax. If we had not already gathered these datasets during our initial keyword search on Google Scholar, we collected them at this stage. We also maintained the same publication year criterion, considering only datasets published between 2016 and 2024. Consequently, Credbank, which was published in 2015 in ICWSM'15, was excluded. BS Detector was no longer publicly available.

\subsection{Claims Datasets}

A summary description of each of the claims datasets can be found in Appendix~\ref{app:datadetails}. Of these datasets, 12 consist of claims scraped from fact-checking or reliable websites, another 12 consist of tweets, and the remaining 11 comprise claims drawn from Twitter, the internet, social media, or news websites. There is variation in the topics of these datasets, but most focus on areas with significant societal impact where misinformation is prevalent and potentially harmful. For example, 16 of the datasets focus on health, vaccination, and COVID-19; 3 focus on political issues; 1 on environmental issues, and the rest covers various subjects, from culture, sport, the economy and so on. Unfortunately, a significant limitation of much of this data is the absence of information regarding the date the claim was made or fact-checked. This can potentially impact the accuracy of labeling, given that certain claims may have been true or false at the time they were made. In addition, this limitation affects the scope of our temporal leakage analysis. Consequently, scholars, and practitioners alike should be cautious when using these data.

\subsection{Paragraph Datasets}

A summary description of each of the paragraph datasets can be found in Appendix~\ref{app:paragraphdetails}. Among the 9 datasets, 8 of the datasets focused on general political misinformation and 1 of the datasets focused on pop culture misinformation. A notable difference in the labeling approaches between claims and paragraphs is that paragraphs tend to rely more on crowd-sourced and source-based labeling approaches, whereas claims datasets tend to rely more heavily on human experts. This makes sense to the extent that relying on human experts is unfeasible for large quantities of text. Paragraph datasets are defined as datasets that strictly contain data types that are either articles, Facebook posts, information from Wikipedia, news articles, or magazine articles. Unfortunately, a significant limitation is that there are many datasets where there is a mixture of claim data types and paragraph data types. There are 39 paragraphs that include paragraph data types, but only 9 of them purely contain paragraph data types without any claim data types. Since the other 30 hybrid datasets give no method of distinguishing between claim data types and paragraph data types in the datasets themselves, we exclude these hybrid cases from the set of paragraph datasets. This illustrates that there is a need for more pure paragraph datasets in the literature.

\subsection{Labeling Approach}

Harmonizing labels across different datasets is a crucial step to ensure comparability of results and robustness of analyses. Since each study employs its own criteria for classifying veracity (see Appendix \ref{app:datadetails}), in this survey, we use the original labels from the studies to create a more consistent categorical variable across datasets. Specifically, we classify content as true, false, mixed or unknown. Information that is mostly true is classified as \textit{true}, while content that is mostly false is classified as \textit{false}. Finally, ambiguous claims, such as those partially true or false, were classified as \textit{mixed}, while claims that were unproven, unrelated or contained no information about their veracity were coded as \textit{unknown}. The percent of true and false claims in each dataset using this coding scheme is shown in Table~\ref{tab:factors}. Moreover, the original annotation method for each dataset, along with its advantages and disadvantages, is detailed in Appendix~\ref{app:labelingapproach}.

%% file: 020relatedwork.tex
\subsection{Expanded Related Work}
\label{sec:background}

In recent years, the scientific community has shown a growing interest in fake news detection to mitigate the spread of misinformation, defined as ``false or misleading information'' \citep{lazer2018science}. Within this evolving field, several surveys have emerged to offer comprehensive reviews and standardized evaluations. A pioneering effort by \citet{shu2017fake} provided an early framework, defining fake news, detailing its characteristics, and summarizing detection techniques from a data mining perspective. Subsequent surveys, such as \citet{oshikawa2018survey} and \citet{zhou2020survey}, have explored alternative methodologies, focusing respectively on natural language processing (NLP) methods and interdisciplinary perspectives. However, while these surveys and others \citep{bondielli2019survey, gravanis2019behind} are valuable for providing a comprehensive overview of the state-of-art in fake news detection, they pay limited attention to existing datasets. Indeed, even if some emphasize the challenges of data collection or stress the importance of dataset quality, these surveys usually provide only superficial coverage of existing datasets, overlooking their specific content, details, and characteristics. Assessing the quality of these misinformation datasets is critical because they are often used to train and test models for misinformation detection and related tasks. A lack of quality data in this context implies that biases and erroneous conclusions could be introduced both in the development stage and in the validation process of these systems.

This gap has thus spurred the emergence of additional surveys dedicated to addressing these dataset-centric nuances, which can be categorized into two types. The first one focuses on categorizing existing datasets to guide the research community in their selection. For example, \citet{d2021fake} surveyed 27 datasets based on eleven characteristics (e.g., application purpose, type of misinformation, language, size, news content type, etc.) and compared these quantitatively. Another example, \citet{sharma2019combating} summarized the characteristic features of 23 existing datasets, providing a clearer picture of those available to the public. However, these surveys have an important drawback; they often lack in-depth analysis. In fact, only descriptive characteristics are listed, thus neglecting key characteristics of their quality and effectiveness for future research. This is also the case for \citet{ali2022fake} and \citet{patra2022understanding}, which describe 26 and 7 datasets, respectively.

The second type of survey focuses on analyzing the quality, performance, and limitations of datasets. For instance, \citet{abdali2024multi} examines 10 datasets to identify some of these weaknesses and strengths. However, a broad approach is used to outline biases in this study, which fails to detail the specifics of each dataset, leaving researchers uncertain about their individual quality. Another example, \citet{hamed2023review} highlight the limitations of 20 articles using publicly available datasets. While this approach provides a good overview of literature trends, a grey area remains regarding whether the errors in these 20 articles stem primarily from methodology or dataset issues. We also find the work of \citet{pelrine2021surprising}, who evaluate the quality of six datasets, focusing on their potential spurious correlations with temporal information. \citet{wu2022probing} expands the analysis of the spurious correlations issue to those induced by event-based collection, dataset merges, and labeling bias, using the Twitter15, Twitter16, and PHEME datasets. However, in both of these studies, the limited number of datasets analyzed fails to provide a comprehensive view of the diverse landscape of available datasets in this field. Similarly, \citet{pelrine2023towards} highlight issues of ambiguous claims in the LIAR dataset, but does not expand their analysis beyond that one and their own LIAR-New dataset.

In short, existing works often only briefly discuss the structure and the content of datasets when addressing data issues, frequently lacking detailed analysis or focusing on a limited number of cases. To overcome this problem, we present one of the most comprehensive surveys of misinformation datasets to date by analyzing their overall content and potential effectiveness in detecting false information.

%% file: tables/label_quality.tex
\begin{table*}[h!]
\caption{Labeling approach and label distribution for 36 claim datasets and 9 paragraph datasets (subset of Table \ref{tab:datasets-summary}). 
}
\label{tab:factors}
\centering
\fontsize{9}{11}\selectfont
\resizebox{\linewidth}{!}{
\begin{tabular}[t]{llrrrrr}
\toprule
Dataset & Labeling approach & True (\%) & False (\%) & Mixed (\%) & Unknown (\%) \\
\midrule
AntiVax & Human expert & 38.15 & 61.85 & - & - \\
Check-COVID & Human expert & 37.92 & 37.16 & - & 24.92 \\
ClaimsKG & Human expert & 17.23 & 63.06 & 12.32 & 7.39 \\ 
Climate-Fever & Human expert & 42.61 & 16.48 & 10.03 & 30.88 \\
CMU-MisCOV19 & Human (N.S.) & 7.39 & 70.63 & - & 21.98 \\
CoAID & Source-based categorization (T) \& Human expert (F) & 93.47 & 6.53 & - & -\\
Counter-covid-19-misinformation & Human (N.S.) & 1.08 & 1.09 & - & 97.83 \\
COVID-19-Rumor & Human expert & 26.16 & 51.27 & - & 22.57 \\
Covid-19-disinformation & Crowd-sourced & -  & -  & - & 100 \\
COVID-Fact &  Human-expert (T) \& Algorithm-generated creation (F) & 31.72 & 68.28 & - & - \\
Covid-vaccine-misinfo-MIC & Crowd-sourced & - & - & - & 100 \\
DeFaktS & Human expert & 11.12 & 7.78 & - & 81.1 \\
ESOC Covid-19 & Human expert & - & 99.52 & - & 0.48\\
FakeCovid & Human expert & 0.85 & 94.12 & 2.74 & 2.28 \\
FaVIQ & Algorithm \& Validation by human & 49.77 & 50.23 & - & - \\
FEVER & Crowd-sourced & 46.75 & 19.65 & - & 33.6\\
FEVEROUS & Algorithm & 57.77 & 38.77 & - & 3.47 \\
FibVID & Human expert & 23.76 & 76.24 & - & - \\
HoVer & Crowd-sourced & 49.76 & 34.95 & - & 15.28\\
IFND & Human expert & 66.64 & 33.36 & - & - \\
LIAR & Human expert & 52.29 & 27.95 & 19.7 & 0.06 \\
LIAR-New & Human expert & 19.62 & 72.87 & 7.51 & \\
MediaEval & Source-based categorization & 44.31 & 51.92 & - & 3.77\\
MM-COVID & Human expert & 71.74 & 28.26 & - & - \\
MultiClaim & Human expert & - & 57.66 & 16.49 & 25.85 \\
NLP4IF-2021 & Crowd-sourced & 64.31 & 3.28 & - & 32.41\\
PHEME & Human expert and non-expert & 33.77 & 66.23 & - & - \\
PubHealthTab & Human expert & 52.47 & 23.79 & - & 23.74\\
Rumors & Algorithm & 11.25 & 58.25 & 6.47 & 24.03 \\
Snopes Fact-news & Human expert & 16.07 & 65.41 & 10.95 & 7.58 \\
TruthSeeker2023 & Crowd-sourced & 51.36 & 48.64 & - & - \\
Twitter15 & Human expert & 50.07 & 24.83 & - & 25.1\\
Twitter16 & Human expert & 50.37 & 25.06 & - & 24.57\\
Verite & Human expert & 33.77 & 66.23 & - & - \\
WICO & Human expert & 68.32 & 31.68 & - & - \\
X-Fact & Human expert & 30.26 & 59.64 & 6.33 & 3.78 \\
\midrule
BanFakeNews & crowd-sourced & 95.56 & 4.44 & - & - \\
Benjamin Political News Dataset & source-based & 39.26 & 60.74 & - & - \\
Celebrity & source-based & 50.00 & 50.00 & - & - \\
CT-FAN & source-based & 26.97 & 64.78 & - & 8.25 \\
FA-KES & crowd-sourced & 52.99 & 47.01 & - & - \\
FakeNewsCorpus & source-based & 1.20 & 79.20 & - & 19.60 \\
FakeNewsAMT & crowd-sourced & 50.00 & 50.00 & - & - \\
ISOT Fake News & source-based & 47.70 & 52.30 & - & - \\
TI-CNN & source-based & 40.34 & 59.66 & - & - 
\\
\bottomrule
\multicolumn{3}{l}{(T) indicates the method used to establish true claims} \\
\multicolumn{3}{l}{(F) indicates the method used to determine false claims} \\
\multicolumn{3}{l}{(N.S.) indicates that expertise is not specified} \\
\end{tabular}}
\end{table*}

%% file: tables/keywords_correlations.tex
\begin{table*}[htbp]
\caption{Keywords correlations evaluation. A high predictivity score which far exceeds its corresponding baseline, means that the keywords provide an unrealistically strong prediction. All numbers are \% macro F1.}
\label{tab:keywords_correlations}
\centering
\resizebox{\textwidth}{!}{
    \begin{tabular}[t]{lcccc}
    \toprule
    Dataset & Keywords Predictivity T-F & Baseline T-F & Keywords Predictivity T-F-M & Baseline T-F-M \\
    \midrule
    Check-COVID & 44.9 & 47.3 & - & - \\
    ClaimsKG & 44.0 & 50.1 & 27.0 & 33.6 \\
    Climate-Fever & 44.9 & 50.2 & 26.7 & 33.4 \\
    CoAID & 60.9 & 50.2 & - & - \\
    COVID-19-Rumor & 48.9 & 50.7 & - & - \\
    COVID-FACT & 40.6 & 52.3 & - & - \\
    DeFaktS & 37.1 & 49.9 & - & - \\
    FakeCovid & 49.9 & 49.9 & 32.7 & 32.8 \\
    FaVIQ & 34.8 & 50.4 & - & - \\
    FEVER & 41.3 & 50.0 & - & - \\
    FEVEROUS & 37.4 & 49.5 & - & - \\
    FibVID & 53.3 & 49.6 & - & - \\
    HoVer & 38.1 & 49.5 & - & - \\
    IFND & 82.2 & 50.0 & - & - \\
    LIAR & 39.5 & 49.3 & 22.9 & 34.5 \\
    LIAR-New & 48.1 & 50.0 & 31.4 & 31.9 \\
    MM-COVID & 77.1 & 51.2 & - & - \\
    NLP4IF-2021 & 48.8 & 52.4 & - & - \\
    PubHealthTab & 42.6 & 49.5 & - & - \\
    Rumors & 45.6 & 51.3 & 29.0 & 34.7 \\
    Snopes Fact-news & 44.1 & 51.7 & 27.6 & 31.8 \\
    TruthSeeker2023 & 66.8 & 50.0 & - & - \\
    Twitter15 & 62.2 & 48.9 & - & - \\
    Twitter16 & 66.4 & 43.3 & - & - \\
    Verite & 39.9 & 49.0 & - & - \\
    X-Fact & 45.8 & 49.8 & 29.2 & 33.5 \\
    \midrule
    BanFakeNews & 91.8 \\
    BenjaminPoliticalNews & 72.2 \\
    Celebrity & 64.0 \\
    CT-FAN & 62.2 \\
    FA-KES & 51.5 \\
    FakeNewsAMT & 45.7 \\
    ISOT Fake News & 91.8 \\
    TI-CNN & 89.1 \\
    \bottomrule
    \end{tabular}
}
\vspace{-1mm}
\end{table*}

%% file: tables/keywords.tex
\begin{table*}[h!]
\centering
\caption{Identification of spuriously predictive keywords.}
\label{tab:keywords}
\centering
\small
\resizebox{\linewidth}{!}{
\begin{tabular}{lcccccccccccc}
\toprule
\multicolumn{4}{c}{IFND} & \multicolumn{3}{c}{MM-COVID} & \multicolumn{3}{c}{Truthseeker2023} & \multicolumn{3}{c}{Twitter16} \\
\cmidrule(lr){2-4} \cmidrule(lr){5-7} \cmidrule(lr){8-10} \cmidrule(lr){11-13} 
 & Fact & Viral & Court & Clinic & Government & India & Biden & Harris & Trump  & Steve & Potus & Poll \\ 
\midrule
True & 17 (0.18\%) & 59 (2.06\%) &  937 (91.95\%) & 134 (100\%) & 97 (98\%) & 142 (99.3\%) & 44 (0.41\%) & 0 (0\%) & 6,229 (63.05\%) & 0 (0\%) & 11 (100\%)& 0 (0\%)\\ 
False & 9367 (99.82\%) & 2812 (97.94\%) & 82 (8.05\%) & 0 (0\%) & 2 (2\%) & 1 (0.07\%) & 10,620 (99.59\%) & 2,584 (100\%) & 3,651 (36.95\%) & 15 (100\%)& 0 (0\%) & 11 (100\%)\\ 
Other & 0 (0\%) & 0 (0\%) & 0 (0\%) &  0 (0\%) & 0 (0\%) & 0 (0\%) & 0 (0\%) & 0 (0\%) & 0 (0\%)& 0 (0\%) & 0 (0\%)& 0 (0\%)\\
\bottomrule
\end{tabular}
}
\end{table*}

%% file: tables/temporal.tex
\begin{table}[htbp]
\caption{Temporal correlations evaluation. A high score here means time---and information correlated with it---is unrealistically predictive.}
\centering
\fontsize{7}{9}\selectfont
\begin{tabular}[t]{lcc}
\toprule
Dataset & Evaluation Type & Temporal Predictivity (\% F1) \\
\midrule
ClaimsKG & Date & 62.3\\
CoAID & Date & 48.3\\
DeFaktS & Date & 37.4 \\
FakeCovid & Date & 49.9 \\
FibVID & Date & 62.2\\
LIAR-New & Date & 53.7\\
Rumors & Date & 74.2 \\
X-Fact & Date & 61.5 \\
AntiVax & TweetID & 46.8 \\
CMU-MisCOV19 & TweetID & 45.6 \\
Covid-19-disinformation & TweetID & 46.5 \\
MediaEval & TweetID & 72.2 \\
Twitter15 & TweetID & 85.6 \\
Twitter16 & TweetID & 95.9 \\
WICO & TweetID & 40.7 \\
\midrule
BanFakeNews & Dates & 98.3 \\
FA-KES & Dates & 50.1 \\
IsotFakeNews & Dates & 87.7 \\
TI-CNN & Dates & 78.1 \\
\bottomrule
\end{tabular}
\label{tab:temporal}
\vspace{-1mm}
\end{table}